\pdfoutput=1

\documentclass[usenatbib]{mn2e}

\usepackage{amssymb}
\usepackage{amsmath}
\usepackage{graphicx}
\usepackage{numprint}
\usepackage{subfigure}
\usepackage{color}

\def\<{\left \langle}
\def\>{\right \rangle}
\def\[{\left\lbrack}
\def\]{\right\rbrack}
\def\({\left(}
\def\){\right)}
\newcommand{\be}{\begin{equation}}
\newcommand{\ee}{\end{equation}}
\newcommand{\ba}{\begin{eqnarray}}
\newcommand{\ea}{\end{eqnarray}}
\newcommand{\prt}{{\partial}}

\newcommand{\red}{{\mbox{\tiny red}}}

\newcommand{\mx}{{\mbox{\tiny max}}}

\title{Modified gravity models and the central cusp of dark matter halos in galaxies}
\author[D.C. Rodrigues et al]{Davi C. Rodrigues,$^{1}$ \thanks{E-mail:davi.rodrigues@cosmo-ufes.org}
Paulo L. de Oliveira,$^{1}$
J\'ulio C. Fabris,$^{1}$ \newauthor and 
Gianfranco Gentile$^{2,3}$ \\ 
$^{1}$Departamento de F\'{\i}sica, Universidade Federal do Esp\'{\i}rito Santo, Av. F.Ferrari, 514, 29075-910, Vit\'oria, Brazil\\
$^{2}$Sterrenkundig Observatorium, Universiteit Gent, Krijgslaan 281 S9, B-9000 Gent, Belgium \\
$^{3}$Department of Physics and Astrophysics, Vrije Universiteit Brussel, Pleinlaan 2, 1050 Brussels, Belgium
}
\begin{document}

\date{}

\pagerange{\pageref{firstpage}--\pageref{lastpage}} 

\maketitle

\label{firstpage}

\begin{abstract} 
\noindent

The N-body dark matter (DM) simulations point that DM density profiles, e.g. the NFW halo, should be cuspy in its center, but observations disfavour this kind of DM profile. Here we consider wether the observed rotation curves “close” to the galactic centre can favour modified gravity models in comparison to the NFW halo, and how to quantify such difference. Two explicit modified gravity models are considered, MOND and a more recent approach called RGGR (in reference to Renormalization Group effects in General Relativity). It is also the purpose of this work to significantly extend the sample on which RGGR has been tested in comparison to other approaches. 

By analysing 62 galaxies from five samples, we find that: i) there is a radius, given by half the disk scale length, below which RGGR and MOND can match the data about as well or better than NFW, albeit the formers have fewer free parameters; ii) considering the complete rotation curve data, RGGR could achieve fits with better agreement than MOND, and almost as good as a NFW halo with two free parameters (NFW and RGGR have respectively two and one more free parameters than MOND).

\end{abstract}


\section{Introduction}

Currently, the main source of issues on the consistency of the standard cosmological model ($\Lambda$CDM) with observational data resides at galactic scales. Among other issues, the $\Lambda$CDM model has difficulties on explaining certain aspects and trends inferred from disk galaxies observations \citep[e.g.,][]{Moore:1999gc, Gentile:2004tb, 2005ApJ...634L.145G, 2009NJPh...11j5029P, deBlok:2009sp, 2011ApJ...736L...2O,2012AJ....143...40M,Adams:2014bda}. One of the main issues of the $\Lambda$CDM model resides on the prediction, from numerical simulations, of a universal cuspy dark matter profile\footnote{Although formally the Einasto profile is not a cuspy one, since its density does not diverge at the centre, the density increase towards the centre is sharp enough to lead in practice to the same issues of the NFW profile \citep{Chemin:2011mf}.} \citep{Navarro:1996gj, Moore:1999gc, 2010MNRAS.402...21N}. Observations on disk galaxies rotation curves (particular on low-surface-brightness (LSB) and some dwarf galaxies) do not favour the existence of such large DM density in the centre of galaxies \citep[for a review, see][]{deBlok:2009sp}. Diverse baryonic mechanisms that could  significantly alleviate the cusp are being studied \citep[e.g.,][]{DelPopolo:2009df,DelPopolo:2011cj,deSouza:2011pb,Pontzen:2011ty,Governato:2012fa, DelPopolo:2014yta}. 

The present work also deals with two approaches outside the standard cosmological model. The first one considers the possibility of the running of the gravitational coupling $G$ in the Renormalization Group context applied to General Relativity, the RGGR model \citep{Rodrigues:2009vf,Farina:2011me,Rodrigues:2012qm}. The second one is the MOND model \citep{1983ApJ...270..365M,1983ApJ...270..371M, Famaey:2011kh}, which proposes the introduction of a fundamental acceleration scale, such that systems with sufficiently small accelerations have dynamics that deviate from Newtonian gravity. For both cases, the gravitational dynamics is modified from the Newtonian one, while the (non-baryonic) dark matter content inside galaxies is set to be either zero or negligible.  Hence, by their definition, these models have no ``cuspy'' dark matter halos. On the other hand, the phenomenological problem with the cuspy halos is not the cusp by itself, but its consequences to the rotation curve. Can the rotation curve data 
``close'' to the centre favour these alternatives over the NFW halo? If one is to consider the NFW halo with a single free parameter, the other being constrained by cosmological arguments, there are some examples where MOND, for instance, can fit the data close to the galactic centre better than the 1-parameter NFW halo \citep{Gentile:2004tb}. An important drawback issue of this approach is that cosmology implies only a correlation of the NFW halo parameters, not that one parameter can be written as a function of the other, hence the fact that a single or a few galaxies are not compatible with that NFW correlation is not by itself in contradiction with the simulations. On the other hand, even the NFW halo with two free parameters is  not favoured when compared to cored halos, and the origin of the discrepancy comes from the central part of the galaxies \citep{deBlok:2002tg}.  

Concerning our MOND analysis in this work, our main purpose is not to make RGGR a MOND competitor, since their principles are quite different. While MOND was mainly developed as a way to avoid the introduction of non-baryonic dark matter, and also to provide a theoretical basis for the Tully-Fisher relation, the motivation of RGGR came from the possibility of measuring nontrivial Renormalization Group effects at large scales. Hence the comparisons done in this work between these models work as a benchmark: since RGGR has an additional free parameter ($\bar \nu$), if its fits were worse than those of MOND, then the former would be a rather weak phenomenological model. 

\bigskip

\section{The models} \label{sec:models}

Here we briefly review the three models that are analysed in this work.  Since the RGGR model is the most recent and the less well known of the three, an extended review is presented for it.

\subsection{Modified Newtonian Dynamics (MOND)}
 
Modified Newtonian Dynamics (MOND) is currently the most influent and perhaps the most successful approach to explain the missing matter problem in galaxies without dark matter. MOND was proposed by \cite{1983ApJ...270..371M,1983ApJ...270..365M}, and it has passed diverse tests at galactic scales since its proposal \citep[see][for a recent review]{Famaey:2011kh}. It models a non-Newtonian gravity in which there is a fundamental acceleration scale (fixed by the constant $a_0$), and, in particular, it can both lead to reasonable rotation curves without the need of dark matter and also provide a basis for the Tully-Fisher relation \citep[e.g.,][]{2012AJ....143...40M}.

Galaxy rotation curves within MOND are obtained from the following simple relation between the Newtonian acceleration $\boldsymbol{a}_N$ and the observed one $\boldsymbol{a}$,
\be
	\boldsymbol{a} ~ \mu \left( \frac{a}{ a_0} \right) = \boldsymbol{a}_N,
\ee
where $a_0 = 1.2 \times 10^{-8}$ cm/$s^2$ \citep{Sanders:2002pf}, $\boldsymbol{a}_N = - \boldsymbol{\nabla} \Phi_N$ and $\Phi_N$ is the standard Newtonian potential derived for the system mass density. The interpolation function $\mu$ is defined such that $\mu (x \ll 1) = x$, and $\mu (x \gg 1) = 1$. This property alone allows for diverse tests of MOND, nevertheless in order to model galaxy RCs, an explicit form of $\mu$ is necessary. \cite{1983ApJ...270..371M}  proposed $\mu(x) = x / \sqrt{ 1 + x^2}$, which is commonly called the original interpolation function, and this is the only one that is studied here. Considering the THINGS sample, this is not the known option that most favours MOND \citep{Gentile:2010xt}, but it is currently unclear whether the latter result also applies to the other samples here studied. Nevertheless, the differences between these choices for the interpolation function do not lead to expressive changes in the results, in most of the cases. Also, since there is no justification for neither of the interpolation functions, some samples may fit better with a given $\mu(x)$, while others may adjust better with another one.

\subsection{The Navarro-Frenk-White (NFW) dark matter halo}
From N-body cold dark matter (CDM) simulations, in the context of $\Lambda$CDM,  \cite{1996ApJ...462..563N} derived the following universal CDM density profile \citep{Navarro:1996gj},
\be
  \rho_{\mbox{\tiny NFW}}(r) = \frac{\rho_s}{\frac{r}{r_s} \( 1  + \frac{r}{r_s}\)^2 },
  \label{nfwprofile}
\ee
where $\rho_s$ and $r_s$ are constants that caracterize a given halo. The NFW halo is a 2-parameter halo, likewise other commonly cited DM profiles, like the (pseudo-)isothermal \citep{1991MNRAS.249..523B} or the Burkert \citep{1995ApJ...447L..25B} ones. For instance, the Burkert halo is given by
\be
  \rho_{\mbox{\tiny Burkert}} (r)= \frac{\rho_0}{\(1 + \frac{r}{r_c}\) \(1 + \frac{r^2}{r_c^2}\)},
\ee
where $\rho_0$ and $r_c$ (the core radius) are constants.  Contrary to the latter halos, the NFW profile has a cusp at the centre, i.e. $ \rho_{\mbox{\tiny NFW}}(r \ll r_s) \approx (r/r_s)^{-1} \rho_s$, while the other cited halos have a core, e.g. $\rho_{\mbox{\tiny Burkert}} (r \ll r_c) \approx \rho_0$. Also contrarily to the NFW halo, the cored halos are known to fit well diverse data, but their derivation in a cosmological context is unclear. One possibility is that the core-like behaviour may be the result of diverse baryonic effects \citep[e.g.,][]{DelPopolo:2014yta}.

The CDM simulations not only suggest the existence of the universal profile (\ref{nfwprofile}), but also pose limits and a correlation between $\rho_s$ and $r_s$. These relations are commonly expressed through the parametrization with $M_{200}$ (or the virial mass $M_{\mbox{\tiny vir}}$) and the concentration $c$. There are different (and quite similar) parametrizations that are commonly adopted for the the NFW halo \citep{0521857937}.

Let $M_{\mbox{\tiny NFW}}(r)$ be the total DM mass inside the radius $r$. The $r_{200}$ is implicitly defined by 
\be
  M_{\mbox{\tiny NFW}}(r_{200}) = 200 \, \frac{4}{3}\pi  r_{200}^3  \, \rho_{\mbox{\tiny crit}},
  \label{r200}
\ee
where $\rho_{\mbox{\tiny crit}}$ is the cosmological critical density. On the other hand, by integrating (\ref{nfwprofile}), 
\be
  M_{\mbox{\tiny NFW}}(r) = 4 \pi \rho_s r_s^3 \[\ln \(1 + \frac{r}{r_s}\) - \frac{\frac{r}{rs}}{1 + \frac{r}{r_s}}  \],
  \label{MrNFW}
\ee
hence
\be
  M_{200} \equiv  M_{\mbox{\tiny NFW}}(r_{200}) = 4 \pi \rho_s r_s^3 \[\ln \(1 + c\) - \frac{c}{1 + c}  \],
  \label{M200}
\ee
where the concentration is defined as $c = r_{200}/{r_s}$. 

The galaxies can be fitted considering $r_s$ and $\rho_s$ as free parameters, then, from the knowledge of these parameters and $\rho_{\mbox{\tiny crit}}$, $r_{200}$ can be (numerically) derived from the combination of eqs. (\ref{r200},\ref{M200}). The concentration $c$ is found from its definition, and hence eq. (\ref{M200}) fixes $M_{200}$.

Simulations also find a correlation between $M_{200}$ and $c$. In this work we propose to test and compare the NFW profile with the least amount of hypothesis on such relations. Since the mentioned relations are statistical in nature, additional issues on bias on the galaxy sample selection would naturally emerge. Besides considering the NFW profile without constraints ($r_s$ and $\rho_s$ free), we also consider the effects of two constraints, see Sec. \ref{sec:variations}.

\subsection{Renormalization group effects on general relativity (RGGR)}

The Renormalization Group (RG) is a well established part of Quantum Field Theory (QFT) in flat space-time (e.g.,  \citealt{Cheng:1985bj, Weinberg:1996kr}). It is well known that the validity of QFT perturbative expressions depend not on the bare couplings values (the coupling constants that appear in the action before any renormalization procedure), but  on the values of the effective couplings, and these values depend both on the bare values and on the energy scale (which also may be interpreted as a distance scale) at which the measurements are performed. 

The dependency between the coupling constant and the energy scale is usually expressed by a beta-function, $\beta_i = \mu \frac{d g_i }{d \mu}$, where $g_i$ are the effective coupling constants and $\mu$ the RG scale. Therefore, if the beta-functions of a model are known,  the functions $g_i(\mu)$ can  be determined.  For instance, in quantum electrodynamics (QED), it is possible to determine the following beta function that describes the QED coupling ($e$) running  (using $c=\hbar =1$),  $ \beta_{QED}= \frac{1}{12 \pi^2} e^3 + O(e^5)$. Hence the effective coupling $e$, which is the electron charge absolute value, increases with the scale $\mu$ (up to some point beyond which the perturbative picture breaks). In the context of scattering experiments with electrons, this means that the effective charge increases with their momenta.  For instance, at $\sim 80$ GeV the value of the fine structure constant $\alpha$ changes from its low energy value $\sim 1/137$ to $\sim 1/128$; for recent results see for instance 
\citep{Acciarri:2000rx, Jegerlehner:2008rs}.

For quantum electrodynamics (QED), one can prove that the corresponding beta function goes to zero in the limit of small energies, such that one has classical electrodynamics in the low energy limit \citep{Appelquist:1974tg}, see also \citep{Goncalves:2009sk} for a recent alternative derivation.

In the case of gravity, as given by the Einstein-Hilbert action, the situation is less clear. The Einstein-Hilbert action is a nonrenormalizable theory  (in the perturbative sense) and currently there is no established complete quantum gravity theory. Nevertheless, in spite on the knowledge or existence of a quantum gravity theory, it is possible to consider the quantization of matter fields in a classical curved background, e.g. \citep{Birrell:1982ix, Buchbinder:1992rb, 0521877873}. The Renormalization Group (RG) formalism can be extended to the context of QFT in curved space-time \citep{Nelson:1982kt} \citep[for a review, see ][]{Shapiro:2008sf}, and it can affect the two dimensionful  constants that characterise the scales of the gravitational effects: the gravitational coupling $G$ and the cosmological constant $\Lambda$. This feature happens in both this and other approaches, e.g.  \citep{Julve:1978xn, Salam:1978fd, Fradkin:1981iu, Nelson:1982kt,Goldman:1992qs, Shapiro:1999zt, Bonanno:2001hi, Reuter:2004nx,Niedermaier:2006wt, Shapiro:2009dh,Weinberg:2009wa, Sola:2013fka}.

\bigskip

In order to derive the large scale gravitational dynamics within the RGGR approach, one needs three  ingredients, which will be motivated and explained shortly: i) the RG-improved action, ii) the beta-function for $G$, and iii) the relation of the scale $\mu$ with observational quantities. These items  are sufficient for fixing the running of $\Lambda$. The main difference between RGGR and similar approaches, based on the RG, relies on the third item above.

Contrary to the case of QED, there is no proof that the running of these constants ($\Lambda$ and $G$), as induced by the RG, become null at very low energies or large scales \citep{Gorbar:2002pw, Shapiro:2009dh}. Moreover, it is unfeasible  to determine the beta function associated with either $G$ or $\Lambda$ entirely from first principles and without additional hypothesis. However, if the gravitational coupling $G$ runs, a natural beta function for it that appeared many times in the literature within different contexts is given by \citep{Reuter:2004nx, Shapiro:2004ch, Farina:2011me},
\be
	\mu \frac{dG^{-1}}{d \mu} =  2 \nu \frac{M_{\mbox{\tiny Planck}}}{c^2 \hbar} = 2 \nu G_0^{-1},
\ee
where $\mu$ is the RG scale parameter (which will be specified) and $\nu$ is a dimensionless parameter that characterises the strength of the RG effects, whose precise value is expected to be found from observations. In particular, if $\nu=0$ there is no running and one recovers standard General Relativity. The above RG equation can be trivially integrated and it yields,
\be
	G(\mu) = \frac{G_0}{1 + \nu \ln \( \frac{\mu^2}{\mu_0^2}\)} \approx G_0 \[ 1 - \nu \ln \( \frac{\mu^2}{\mu_0^2}\)\].
	\label{GmuLog}
\ee
In this work,  only the expressions up to first order on $\nu$ will be necessary. 

From a phenomenological perspective, eq.(\ref{GmuLog}) is  a natural choice for studying possible variations of $G$. Since, considering the many tests that pure General Relativity have passed, if $G$ runs at large scales, it should run ``slowly''.

The action that extends the  Einstein-Hilbert action, by implementing the large scale RG effects, is a functional of the space-time metric $g_{\mu \nu}$ and the Renormalization Group (RG) scale $\mu$. It reads\footnote{On our conventions: We use the signature (- + + +), $dx^0 = c \, dt$, and the Riemann tensor, Ricci tensor and  Ricci scalar are respectively: $R^\rho_{\; \mu \nu \lambda} \equiv \prt_\nu \Gamma_{\mu \lambda}^\rho + \Gamma^\rho_{\nu \alpha } \Gamma^\alpha_{\mu \lambda} - (\mbox{the same with } \nu \leftrightarrow \lambda)$, $R_{\mu \lambda} \equiv R^\nu_{\; \mu \nu \lambda}$ and $R \equiv g^{\mu \nu}R_{\mu \nu}$.} \citep{Reuter:2004nx,  Shapiro:2004ch, Koch:2010nn, Rodrigues:2012qm}
\be
	S[g,\mu] = \frac{c^3}{16 \pi}\int \frac{R - 2 \Lambda(\mu)}{G(\mu)} \sqrt{-g} \, d^4x.
	\label{RGIA}
\ee

The field equations are found by varying the action with respect to either the metric or $\mu$, and they read

\ba
	\label{RGIAeq}
	&& \!\!\!\!\!\!\!\!\!\!\!\!G_{\mu \nu} +  \Lambda \,  g_{\mu \nu} + G \, \square G^{-1} g_{\mu \nu} - G \, \nabla_\mu \nabla_\nu G^{-1} = \frac{8 \pi G}{c^4} \, T_{\mu \nu} \, , \\
	\label{RGIAEM}
	&&\!\!\!\!\!\!\!\!\!\!\!\! \nabla_\nu \( \frac{\Lambda}G \) = \frac 12  R \, \nabla_\nu G^{-1} \, ,
\ea
where an energy-momentum tensor $T_{\mu \nu}$ was added to account for any matter or fields that interact gravitationally.

Likewise on standard General Relativity, the total stress-energy tensor must be conserved. Indeed, from the energy-momentum conservation ($T_{\mu \nu}^{\; \; \; ;\nu} = 0$), the Bianchi identities ($G_{\mu\nu}^{\;\;\; ; \nu} = 0$), eq. (\ref{RGIAeq}) and the property $ \left (  \square \nabla_\nu - \nabla_\nu \square \right ) G^{-1} = R_{\nu \kappa} \nabla^\kappa G^{-1}$, one finds precisely the eq. (\ref{RGIAEM}) \citep{Koch:2010nn, Rodrigues:2012qm}. And, reciprocally, from eqs. (\ref{RGIAeq},\ref{RGIAEM}) one derives $T_{\mu \nu}^{\; \; \; ;\nu} = 0$.

It remains to be settled the relation of $\mu$ with space-time dependent physical quantities. A general solution for this issue need not to be a simple one, see e.g. \citep{Babic:2004ev, Domazet:2010bk}. In the context of stationary, weak field and low velocity systems, all the General Relativity dynamics can be derived from the Newtonian potential ($\Phi_N$), which, for clarity, we state its definition as 
\be
  \nabla^2 \Phi_N \equiv 4 \pi G_0 \rho,
\ee
where $\rho$ is the matter density and  $\Phi_N (r \rightarrow \infty) = 0$. In the same context, the Newtonian potential also is a good proxy on space-time geometry, being better than the Ricci scalar alone. The latter is probably the simplest fully covariant choice for $\mu$, but cannot differentiate a Schwarszchild space from a Minkowski one. Hence, at least the first order the RG correction should be correlated to $\Phi_N$, namely, there should exist $f$ such that
\be
  \frac{\mu}{\mu_0} = f\(\frac{\Phi_N}{\Phi_0}\).
  \label{muf}
\ee

A proposal that has appeared diverse times in the literature is $\mu \propto 1/r$. This proposal has similarities with an identification with the Newtonian potential for large distances, but it faces some critical issues; particularly, if in this context the RG effects are not negligible for large $r$, it will face strong problems to fit the central part of galaxy rotation curves, as shown by \cite{Rodrigues:2009vf}. 

The simplest solution to eq.(\ref{muf}) capable of providing a good Newtonian limit is\footnote{In particular  $\mu \propto ( \partial_r \Phi_N)^\alpha$ and $\mu \propto (\nabla^2 \Phi_N)^\alpha \propto \rho^\alpha $  were shown to be incompatible with  dark matter-like effects in galaxies \citep{Rodrigues:2011cq}.} \citep{Rodrigues:2009vf}
\be
  \frac{\mu}{\mu_0} = \( \frac{\Phi_N}{\Phi_0} \)^\alpha.
  \label{muset}
\ee
The above expression introduces two quantities, $\Phi_0$ and $\alpha$. The first one only fix the value of the Newtonian potential in which $G=G_0$. Since in the RGGR model the variation of $G$ inside galaxies is small (about $10^{-7}$ of its value across $\sim$ 40 kpc), there is no need to be very precise on the value of $\Phi_0$; moreover, up to first order on $\nu$, $\Phi_0$ simply does not appear in the field equations or in the derivative of the effective potential, the latter will be shown shortly (see also \citealt{RGGRspherical}). The apparently simpler ansantz of using eq. (\ref{muset}) with, say, $\alpha =1$ is incompatible with observations, since in this case the standard General Relativity behaviour  could only be recovered for systems with constant (or about constant) Newtonian potentials. The reason for this is that the conditions $\mu = \mu_0$ or $\Phi_N = \Phi_0$ at some space region are not sufficient for recovering the Einstein field equations. The latter conditions only guarantee that $G 
= G_
0$ in that region, but not that $\mathbf{\nabla} G= \mathbf{0}$ in the same region.

\bigskip

The above is sufficient to set the essential dynamics of the model in the specified context. This model was named RGGR, in a reference to Renormalization Group effects in General Relativity. The next step is to compare the model with observations. Formally, three constants were introduced, $\Phi_0$, $\nu$ and $\alpha$. The precise value of the former is irrelevant for the dynamics if the running is small, as previously commented, while the two remaining constants in practice only appear together in the form $\bar \nu \equiv \nu \alpha$, which can be verified by combining eqs. (\ref{GmuLog},\ref{muset}). Hence for all the phenomenological purposes there is only one more parameter, $\bar \nu$.

Considering the issues of the  standard cosmological model in galaxies, and the possibility that such RG running of $G$ may have impact at the scales of galaxies, we have speculated that  these RG effects may have consequences to the abundance and type of dark matter needed to fit the observational data. If such RG effect is real, how much dark matter is needed in order to have a good agreement with the kinematics of galaxies? According to \citep{Rodrigues:2009vf,Rodrigues:2012qm, Rodrigues:2012wk, deOliveira:2012hfa,Rodrigues:2013bd} the answer is none, that is, disk galaxies rotation curves and elliptical galaxies dispersion curves can be explained without dark matter. One of our aims in this work is to better evaluate this issue with a larger sample of disk galaxies (which are the ones that typically can put tighter bounds on the distribution of dark matter or on modified gravity effects). 

In order to apply this model to galaxies,  one needs stationary, weak field and small velocity solutions with  axisymmetric mass distributions. For the particular case of spherical symmetry, the details can be found in \citep{RGGRspherical}. In general, as long as the contribution of the cosmological constant is negligible, one can use a conformal transformation in order to find the effective potential $\Phi$ from the Newtonian one $\Phi_N$ \citep{Shapiro:2004ch, Rodrigues:2009vf}. To be clear, the effective potential is a field such that $\ddot x^i = - \nabla^i \Phi$. It is straightforward to show that, in this context, $g_{00} = -1 - 2\Phi/c^2$ (the proof is the same one of General Relativity, since the form of the geodesic equation is the same of RGGR). The relation between $\Phi$ and $\Phi_N$ is, apart from an additive constant, given by the following time-time component of the conformal transformation,
\be
	1 + \frac{\Phi}{c^2} = \frac{G}{G_0}  \( 1 + \frac{\Phi_N}{c^2}\).
\ee
Hence, from (\ref{GmuLog}) with $\frac{\mu}{\mu_0} = \( \frac{\Phi_N}{\Phi_0}\)^\alpha$ and $\bar \nu = \nu \alpha$,
\be
	\frac{\Phi}{c^2} = \frac{\Phi_N}{c^2} -  \bar \nu \ln \frac{\Phi_N^2}{\Phi_0^2} + O(\bar \nu^2).
	\label{rggrphi}
\ee

From the above, one arrives at the essential relation between circular velocity and acceleration for stationary systems supported by rotation, namely
\be
	V^2 = r \, \partial_r \Phi = V^2_N \( 1 - \frac{\bar \nu c^2}{\Phi_N}\),
  \label{Vrggr}
\ee
where $V_N$ is the Newtonian circular velocity, it satisfies $V_N^2/r = \partial_r \Phi_N $. For $\bar \nu > 0$, one always has $V^2 > V^2_N$, since $\Phi_N <0$.

Contrary to Newtonian gravity, the RGGR effective potential $\Phi$ has a nonlinear dependence on the mass density of the system. Hence, to derive the $\Phi$ corresponding to a complex system (e.g., a galaxy) from a straightforward integration of the point particle solution is not feasible. Similarly, it is not possible to describe a galaxy in General Relativity from some integration of its point particle solutions (which are black holes). Remarkably, within RGGR, given some matter distribution, one can compute $\Phi_N$, and from eq. (\ref{rggrphi}) find the RGGR effective potential $\Phi$. This quasi-linear process, however,  cannot be used to derive the  $\bar \nu$ value  of the system. For instance, if the corresponding  $\bar \nu$ of every star that compose a galaxy is known, the corresponding $\bar \nu$ of the whole stellar contribution would still be unknown. The single expectation on $\bar \nu$ that we know comes from phenomenology: its value should increase from the Solar System to larger or more massive systems, in order  to have impact on dark matter effects without spoiling the success of General Relativity in the Solar System. A Solar System constraint was presented by \cite{Farina:2011me}. In conclusion, since the value of $\bar \nu$ depends on nonlinear effects that {\it a priori} we have no control, $\bar \nu$  is dealt as a free parameter that can change from galaxy to galaxy.  In Sec. \ref{sec:Yresults}, the correlation between baryonic mass and $\bar \nu$ is briefly evaluated.

\subsection{The variations: MOND$_\delta$, NFW$_{13}$, NFW$_{12}$ and RGGR$_\delta$} \label{sec:variations}

Besides the three main models reviewed above, we also consider four variations. 

Since both MOND and RGGR are here fitted without dark matter, both amplify the Newtonian rotation curve in order to match the observed rotation curve, and hence they are more susceptible to changes in the galaxy baryonic model. In particular, changes of the 10\% order in the galaxy distances lead to negligible effects in the NFW fits\footnote{It will change the values of $r_s$, $\rho_s$ and $\Upsilon_*$ accordingly, but apart from the gas influence it will preserve $c$, $M_{200}$, the baryonic mass and $\chi^2$.} (and many other dark matter models), but such changes may lead to significant changes in models without dark matter.

Diverse works on MOND have addressed the issue of distance variations, and for some galaxies small variations on the distance can be the difference between a poor fit to a good one \citep[e.g.,][]{1998ApJ...508..132D, Gentile:2010xt}. The MOND variation in which the distance is not {\it a priori} fixed to be the one stated in the Table \ref{tab:inputdata} is named here MOND$_\delta$. In this variation, the distance to a given galaxy is given by the product $D \delta$, where $D$ is the distance stated in Table \ref{tab:inputdata}, and $\delta$ is a new parameter to be fitted. In order to avoid unrealistic distances, $\delta$ is constrained such that changes in the distance are not greater than 20\%.

Similarly, RGGR$_\delta$ is a RGGR model in which galaxy distances can vary by no more than 20\%. This is the first time that this model is fitted with non-constant distances.

The fits of NFW halos with free $r_s$ and $\rho_s$ are known to lead in some cases to unrealistic high values of $r_s$ and $M_{200}$ for galaxies, and too low $c$ values (e.g., $M_{200} > 10^{14} M_\odot$ and $c < 1$). A form of avoiding this unrealistic behaviour is to impose a strict relation between $M_{200}$ and $c$, nevertheless such relation would be a stronger input than that given by the simulations, since from the simulations perspective it is only possible to guarantee the existence of a certain correlation (with non-null dispersion) between these parameters \citep{0521857937}. In this work we also evaluate the consequences of constraining $M_{200}$ and its impact on the derived rotation curve close to the galactic centre. Two variations of the NFW dark matter halo model are used, NFW$_{13}$ and NFW$_{12}$. In the first case, it is imposed that $M_{200} \leq 10^{13} M_\odot$, which is a natural constraint considering that almost all of the fitted galaxies have baryonic masses lower than $10^{11} M_{\odot}$. For the second case, NFW$_{12}$, the constraint is $M_{200} \leq 10^{12} M_\odot$. Perhaps this is a too strong constraint for the most massive galaxies, like NGC 2841, but it is a rather reasonable constraint for many of the galaxies studied here. In particular, there are some cases of galaxies with baryonic mass of about $10^9 M_\odot$ and whose resulting fit from the NFW model implied $M_{200} > 10^{12} M_\odot$, see Fig. \ref{fig:M200grid}.


\section{The samples and the methods}

The galaxies studied in this work are divided in the samples given in Table \ref{tab:samples}. Observational global parameters of each galaxy are displayed in Table \ref{tab:inputdata}. In Sec. \ref{sec:chi2R} a quantity ($\chi^2_R$) that will be helpful on the detection of radius dependent systematics is introduced.  In a forthcoming publication, we plan to evaluate issues related to the derived stellar mass-to-light ratios.

\subsection{Galaxies and baryonic models} \label{sec:galax&barmodels}
  Some of the galaxies that are studied here have more than one baryonic model. For instance, in the Sample A the galaxy NGC 2403 appears in two versions, either as a bulgeless galaxy (NGC 2403 1D) and as a galaxy with both stellar disk and bulge (NGC 2403 2D). In this case, these two baryonic models come from the same original reference, and we use here the same conventions employed in the original work. Also, some of the galaxies that appear in the Sample C also appear in either the Sample D or in the Sample E, but with different models for the baryonic part. After eliminating all the redundancies, we are left with 53 unique galaxies from a sample of 62 different baryonic models for galaxies. 

Removing the redundancies is not an unambiguous work, since for some cases it is not clear why a particular baryonic model should be preferred than another one. Therefore, we do not eliminate the redundancies and fit every baryonic model as if it were a unique galaxy. In the end we checked the effect of removing the redundancies by adopting different criteria. No strong dependence on whether the redundancies are eliminated or not was found, hence the results are stated without the elimination of the redundancies. From the Table \ref{tab:results}, which presents our results galaxy by galaxy, one can redo our analysis and check quantitative changes in the results by adopting different conventions on the elimination of the duplicate galaxies. Considering the arguments above and in order to avoid longer explanations, different baryonic models of the same galaxy are counted as different galaxies, unless otherwise stated.

\subsection{The samples} \label{sec:samples}
All the galaxy analysed are disk galaxies of diverse types, from dwarfs to giants, LSB or HSB, galaxies with prominent bulge or without one. The baryonic models of these galaxies are not derived or developed in this work, we use the baryonic models that were derived in the main reference of each one of the five samples. For the sake of clarity, we labeled each one of the five samples with a letter (the order of the samples is arbitrary), see Table \ref{tab:samples}.

Sample A is a recent sample of nearby galaxies whose baryonic models are based on measurements done in diverse bands. It includes detailed modelling of the expected stellar mass-to-light ratio $\Upsilon_*$ considering two stellar initial mass functions (IMF), the Kroupa and the (diet-)Salpeter ones \citep{2008AJ....136.2648D}. This sample contains galaxies of diverse types, but it has a significant bias towards ``large'' galaxies (i.e., the more massive and with higher maximum circular velocity). Fits considering a NFW DM halo were done in the main reference of this sample. Here we redo all the NFW fits both to assure that the comparisons between models are done exactly with the same conventions, and to compute the quantities $\chi_R^2$. Moreover, we present results for the models with constraints on $M_{200}$.

Fits for the Sample A assuming MOND without DM were done in \citep{Rodrigues:2009vf, Gentile:2010xt} (the first reference only studies five galaxies of this sample), and are redone here for the same reasons of the NFW model. Considering RGGR, five galaxies of this sample were studied in \citep{Rodrigues:2009vf}, but assuming that the stellar profile was exactly exponential. These galaxies are redone here without this assumption, and the resulting fits present in this work are indeed very similar to those presented in\footnote{This is not expected to be a general result, those five galaxies were selected in part due to the almost exact stellar exponential profile they display.} \citep{Rodrigues:2009vf}.

Sample B is a sample of five galaxies with luminosities in the range $\sim 6 \times 10^8 - 1 \times 10^{10}$ $L_{\odot, I}$ that has been used many times by other works, in particular due to their evidence against the cusp in dark matter profiles. The stellar components of these galaxies are measured in the I-band, and hence it has relatively sharp constraints on the possible values of $\Upsilon_*$. Four galaxies of this sample have stellar components that are exactly exponential. The MOND and RGGR RCs of these four were also studied in \citep{Rodrigues:2009vf}. The main reference of this sample  presents results for both MOND and NFW, but using different assumptions on the fits, and in particular, in the case of the NFW halo, only results for the the 1-parameter NFW halo are presented.

Samples C, D and E share some similarities and have some galaxies in common, see Sec.\ref{sec:galax&barmodels}. All the three are dominated by low mass galaxies (albeit not restricted to), and many of them are commonly classified as LSB galaxies. The stelar components were measured in the R-band, which is not a band as good as the near infrared ones to estimate $\Upsilon_*$, due to the larger dispersions \citep{Bell:2000jt}, but it is better than the B-band. The rotation curves of Sample C were built by using both H$\alpha$ and HI data. Many tests of diverse models have used the galaxies of the samples C and D. In particular, in the main reference of  the Sample C, there is a comparison of  the 2-parameter NFW halo fits with the (pseudo-)isothermal halo ones, and a statistical preference for the isothermal halo model is derived, the latter due to the RC behaviour in the central part of the galaxy. Results for the fits assuming MOND, for some galaxies of Samples C and D, can be found in \citep{2010ApJ...718..380S}, see also \citep{Brownstein:2005zz,Mannheim:2010xw} for other modified gravity approaches. The main references of the Samples C and D analyse more galaxies than those in our Samples C and D. We only consider the data provided by those references and which include the gas and stellar rotation curves. 

All the 18 dwarf galaxies studied in \citep{Swaters:2011yq} constitute our Sample E. In that reference it is claimed that the maximum disk fits in dwarf galaxies can account for their respective rotation curves close to the galaxy centre, similarly to larger disk galaxies, and contrary to previous claims (see that reference for further details). This discrepancy is a consequence of differences on the RCs steepness close to the galaxy centre that \cite{Swaters:2011yq} derive, which is explained in part due to the beam smearing technique that is used. Hence higher values of $\Upsilon_*$ are possible (considering the RC only). On the issue of dark matter cusps, previous versions of the baryonic models for some of the galaxies of this sample were studied, for instance, in  \citep{Swaters:2002rx}.

Details on stellar mass-to-light ratios and stellar initial mass functions (IMF) will be studied in a future publication.

\begin{table}

\caption{The samples. Col. (2): the number of galaxies here analysed from each of the six samples (galaxies with two different descriptions for the baryonic part count twice).   Col.(3) presents the main references for the observational data.}
\label{tab:samples}
\begin{tabular}{@{}lcc}
\hline
Sample & Fitted galaxies &   Main Refs. \cr
(1) & (2)  & (3) \cr
\hline
A & 18  &\citep{2008AJ....136.2648D}\cr
B & 05  & \citep{Gentile:2004tb} \cr
C & 13  &\citep{deBlok:2002tg}   \cr
D & 08  & \citep{2001AJ....122.2396D}\cr
E & 18 & \citep{Swaters:2011yq}\cr
\hline 
Total & 62 & \hspace*{-0.50in} {\footnotesize different baryonic models for galaxies} \cr
	& 53 & \hspace*{-1.53in}{\footnotesize different galaxies}\cr
\hline 
\end{tabular}
\end{table}


\npdecimalsign{.}
\nprounddigits{1}
\npproductsign{\! \times \!}

\begin{table*}
\caption{Galaxy by galaxy global parameters adopted for the fits here performed. Col.(1): the sample. Col.(2): galaxy name. Col.(3) distance (Mpc). Col.(4): disk luminosity ($L_\odot$), the band depends on the sample (3.6 $\mu$m for sample A, I-band for sample B, R-band for samples C,D and E). Col.(5): same as before for the bulge. Col.(6): the total gas mass ($M_\odot$) (it includes HI and He masses, it may also contain additional elements depending on the sample, see Table \ref{tab:samples} for further details). Col.(7): disk scale length. }
\begin{tabular}{cln{2}{1}n{1}{1}n{1}{1}n{1}{1}n{2}{1}}
 \hline
{S} & {Galaxy} & \multicolumn{1}{c}{Dist.} & \multicolumn{1}{c}{$L_D$} & \multicolumn{1}{c}{$L_B$} & \multicolumn{1}{c}{$M_{\mbox{gas}}$} &{$R_D$} \\ 
{(1)} &    {(2)}   &  \multicolumn{1}{c}{(3)}    &   \multicolumn{1}{c}{ (4)}  &  \multicolumn{1}{c}{ (5)}   &  \multicolumn{1}{c}{(6)}   & \multicolumn{1}{c}{(7)}   \\
\hline
A & DDO 154 & 4.3  & 8.2e7 &\multicolumn{1}{c}{---}&4.626e8 & 1.00  \\
A & NGC 2403 1D & 3.2 & 1.250881424369185e10 &\multicolumn{1}{c}{---}& 3.8213e9 & 1.787 \\ 
A & NGC 2403 2D & 3.2 & 1.1993208750953798e10 & 7.109658646693224e8 & 3.8213e9 & 1.7868513719  \\ 
A & NGC 2841 & 14.1 & 1.481727292085382e11 & 2.9903409898923595e10 & 1.43751e10 & 4.1127457822  \\ 
A & NGC 2903 & 8.9 & 2.3156353190536976e10 & 1.6445862226940248e9 & 6.6047e9 & 2.4412615111  \\ 
A & NGC 2976 & 3.6 & 3.2332352909798594e9 &\multicolumn{1}{c}{---}& 1.463e8 & 1.64 \\ 
A & NGC 3031 & 3.6 & 8.647887136486703e10 & 1.2882495516931322e10 & 4.4243e9 & 2.8626657468 \\ 
A & NGC 3198 1D & 13.8 & 3.1197884386647774e10 &\multicolumn{1}{c}{---}& 1.53729e10 & 3.06 \\ 
A & NGC 3198 2D & 13.8 & 3.522978664080562e10 & 3.95072808647481e9 & 1.53729e10 & 3.1415640817  \\ 
A & NGC 3521 & 10.7 & 1.6852996860443576e11 &\multicolumn{1}{c}{---}& 1.30586e10 & 13  \\ 
A & NGC 3621 & 6.6 & 3.3048213555221043e10 &\multicolumn{1}{c}{---}& 9.5755e9 & 2.5991729916  \\ 
A & NGC 4736 & 4.7 & 2.955693867718834e10 & 1.17892468180085e10 & 3.8213e9 & 1.99 \\ 
A & NGC 5055 & 10.1 & 1.5573022415346594e11 & 1.899360118958219e10 & 1.42119e10 & 2.44 \\ 
A & NGC 6946 & 5.9 & 9.204424852358423e10 & 3.8018939632056127e9 & 5.5733e9 & 2.97 \\ 
A & NGC 7331 & 14.7 & 2.370838439196519e11 & 1.7378008287493763e10 & 1.24587e10 & 5.5 \\ 
A & NGC 7793 & 3.9 & 8.884608720445688e9 &\multicolumn{1}{c}{---}& 1.1557e9 & 1.6393623104\\ 
A & NGC 7793 R & 3.9 & 8.884608720445688e9 &\multicolumn{1}{c}{---}& 1.1557e9 & 1.6393623104\\ 
A & NGC 925 & 9.2 & 1.5742969112011595e10 &\multicolumn{1}{c}{---}& 5.0561e9 & 2.69 \\ 
B & ESO 116-G12 & 15.3 & 4.786e9 &\multicolumn{1}{c}{---}& 1.5121e9  &1.6875 \\
B & ESO 287-G13 & 35.600 & 2.29086765276777e10 &\multicolumn{1}{c}{---}& 1.100032372889039e10 & 3.28125  \\
B & ESO 79-G14 & 30.300 & 1.7378008287493763e10 &\multicolumn{1}{c}{---}& 3.448945921839768e9 & 3.875 \\
B & NGC 1090 & 36.400 & 2.511886431509582e10 &\multicolumn{1}{c}{---}& 8.363002394172916e9 & 3.40625 \\
B & NGC 7339 & 17.800 & 8.317637711026743e9 &\multicolumn{1}{c}{---}& 5.581053948341904e8 & 1.53125 \\
C & F563-1 & 45 & 9.289663867799401e8 &  \multicolumn{1}{c}{---} & 5.306747757497086e9 & 3.5 \\ 
C & UGC 1230 & 51 & 2.558585886905653e9 &  \multicolumn{1}{c}{---} & 8.984504826413485e9 & 4.5 \\ 
C & UGC 3060 & 51 & 3.5645113342624533e8 &  \multicolumn{1}{c}{---} & 1.0951850853928092e9 & 1.3 \\ 
C & UGC 3371 & 12.8 & 7.046930689671452e8 &  \multicolumn{1}{c}{---} & 1.6266566419781063e9 & 3.1 \\ 
C & UGC 3851 & 3.4 & 3.3728730865886927e8 &  \multicolumn{1}{c}{---} & 1.1612620680368497e9 & 1.5 \\ 
C & UGC 4173 & 16.8 & 7.726805850957022e8 &\multicolumn{1}{c}{---}& 3.0084898447070136e9 & 4.5 \\ 
C & UGC 4325 & 10.1 & 1.018591388054119e9 &\multicolumn{1}{c}{---}& 1.0965226214790547e9 & 1.6 \\ 
C & UGC 5005 & 10.1 & 1.6143585568264935e9 &\multicolumn{1}{c}{---}& 5.523176330465866e9 & 4.4 \\ 
C & UGC 5721 & 6.7 & 2.805433637951705e8 &\multicolumn{1}{c}{---}& 1.0226848989958826e9 & 0.5 \\ 
C & UGC 7524 & 3.5 & 1.01859138805412e9 &\multicolumn{1}{c}{---}& 1.27392889394151e9 & 2.3  \\ 
C & UGC 7603 & 6.8 & 3.3728730865886927e8 &\multicolumn{1}{c}{---}& 6.573222573567345e8 & 0.7 \\ 
C & UGC 8837 & 5.1 & 1.1168632477805611e8 &\multicolumn{1}{c}{---}& 2.2919740373352987e8 & 1.2 \\ 
C & UGC 9211 & 12.6 & 1.7701089583174184e8 &\multicolumn{1}{c}{---}& 1.6115617606455114e9 & 1.2 \\ 
D & F563-1 & 45 & 1.1420e9 &\multicolumn{1}{c}{---}& 5.2830e9 & 2.8 \\ 
D & F568-3 & 77 & 2.6894e9 &\multicolumn{1}{c}{---}& 3.9503e9 & 4 \\ 
D & F571-8 & 48 & 4.5463e8 & 9.4986e8 & 6.3068e8 & 2.8 \\ 
D & F579-V1 & 85 & 4.1463e9 &\multicolumn{1}{c}{---}& 3.1460e9 & 5.1 \\ 
D & F583-1 & 32 & 5.4659e8 &\multicolumn{1}{c}{---}& 2.9543e9 & 1.6 \\ 
D & F583-4 & 49 & 7.2054e8 &\multicolumn{1}{c}{---}& 5.4626e8 & 2.7 \\ 
D & UGC 5750 & 56 & 4.0551e9 &\multicolumn{1}{c}{---}& 5.3750e9 & 3.3 \\ 
D & UGC 6614 & 85 & 3.4001e10 & 1.2323e10 & 2.5262e10 & 7.8297 \\ 
E & UGC 11707 & 15.9 & 1.6143585568264935e9 &\multicolumn{1}{c}{---}& 4.919772499749601e9 & 4.3  \\ 
E & UGC 12060 & 15.7 & 8.47227414140598e8 &\multicolumn{1}{c}{---}& 2.48127416623708e9 & 1.76 \\ 
E & UGC 12632 & 6.9 & 8.47227414140598e8 &\multicolumn{1}{c}{---}& 1.1866656462583575e9 & 2.57 \\ 
E & UGC 12732 & 13.2 & 9.289663867799401e8 &\multicolumn{1}{c}{---}& 4.961493345923655e9 & 2.21  \\
E & UGC 3371 & 8 & 7.046930689671452e8 &\multicolumn{1}{c}{---}& 1.6488405386430175e9 & 3.09 \\ 
E & UGC 4325 & 10.1 & 1.018591388054119e9 &\multicolumn{1}{c}{---}& 1.0201216196230191e9 & 1.63  \\ 
E & UGC 4499 & 13 & 7.726805850957022e8 &\multicolumn{1}{c}{---}& 1.606151619914913e9 & 1.49 \\ 
E & UGC 5414 & 10 & 6.426877173170221e8 &\multicolumn{1}{c}{---}& 8.819109700896121e8 & 1.49  \\ 
E & UGC 6446 & 12 & 1.342764961137864e9 &\multicolumn{1}{c}{---}& 1.828490463964886e9 & 1.87 \\ 
E & UGC 731 & 8 & 2.5585858869056532e8 &\multicolumn{1}{c}{---}& 9.804108251359967e8 & 1.65 \\ 
E & UGC 7323 & 8.1 & 2.1281390459827178e9 &\multicolumn{1}{c}{---}& 9.901236750364338e8 & 2.2 \\ 
E & UGC 7399 & 8.4 & 4.05508535448386e8 &\multicolumn{1}{c}{---}& 1.0730061965725577e9 & 0.79 \\ 
E & UGC 7524 & 3.5 & 1.018591388054119e9 &\multicolumn{1}{c}{---}& 1.2604560172599852e9 & 2.58 \\ 
E & UGC 7559 & 3.2 & 1.7701089583174184e7 &\multicolumn{1}{c}{---}& 1.0192530900934324e8 & 0.67 \\ 
E & UGC 7577 & 3.5 & 1.018591388054119e8 &\multicolumn{1}{c}{---}& 1.1527031534190749e8 & 0.84 \\ 
E & UGC 7603 & 6.8 & 3.3728730865886927e8 &\multicolumn{1}{c}{---}& 6.01034232267685e8 & 0.9 \\ 
E & UGC 8490 & 4.9 & 4.87528490103387e8 &\multicolumn{1}{c}{---}& 1.0929609091685169e9 & 0.66 \\ 
E & UGC 9211 & 12.6 & 1.7701089583174184e8 &\multicolumn{1}{c}{---}& 1.4540283823297203e9 & 1.32 \\ 
\hline
\end{tabular}
\label{tab:inputdata}
\end{table*}

\subsection{The quantity $\chi^2_R$ as a tool for detecting systematics at specific galaxy regions} \label{sec:chi2R}

In order to evaluate how well MOND, NFW and RGGR fit rotation curve (RC) data at the galactic inner radii, an important step is to select a quantity capable of quantifying how much each model derived RC deviates from the observational data ``close'' to the galactic centre. It seems rather unnatural to analyse a large sample of galaxies looking for deviations ``close'' to the galactic centre and also within a few kpc, since the extent of the observed RC's may vary  from about 5 kpc to 50 kpc. Hence we analysed rotation curve deviations within a certain fixed amount of disk scale lengths ($R_D$), which consist a natural physical distance scale for disk galaxies in their inner region. This lead to the definition and analysis of three quantities that resemble $\chi^2$: $\chi^2_{\mbox{\tiny 2RD}}$, $\chi^2_{\mbox{\tiny RD}}$, $\chi^2_{\mbox{\tiny  RD/2}}$, which are the values of the derived minimum of $\chi^2$, but only considering the observational data inside two, one or half disk scale lengths. In order to be more specific, first the fit is derived such that the value of $\chi^2$ is minimum, latter the quantity $\chi^2_R$ is computed with
\be
    \chi^2_R = \sum_{i=1}^{i_R} \left( \frac{V_{\mbox{\tiny model}}(R_i) - V_i}{\sigma_i} \right)^2,
\ee
where $V_i$ and $\sigma_i$ are the observed rotation velocity and its error at the radius $R_i$, and  $i_R$ stands for the smallest integer number such that $R \leq R_i$. Therefore, in particular, if $R_\mx$ is the radius of the outermost observed rotation curve data, $ \chi^2_{R_\mx} = \chi^2$. 

Among the 62 galaxies studied, there are 10 that have no observational data inside $R_D/2$, and only one galaxy of the sample that have no observational data inside $R_D$. These numbers do not change when considering the models MOND$_\delta$ and RGGR$_\delta$, since both the position of the observational data (in kpc) and the value of $R_D$ change proportionally with $\delta$. All the medians on $\chi^2_{\mbox{\tiny RD}}$ and $\chi^2_{\mbox{\tiny RD/2}}$ are done after the galaxies with $\chi^2_{\mbox{\tiny RD}}=0$ and $\chi^2_{\mbox{\tiny RD/2}}=0$ are respectively removed (which are  the galaxies with no observational data inside $R_D$ or $R_D/2$).

\subsection{The number of fitted parameters and $\chi^2_\red$} \label{sec:freep}

In the context of galaxy analysis, it is common to employ the reduced chi-squared ($\chi^2_\red$) in order to infer how well a given model  matches the observational data. Considering all the non-linearity involved in the galaxy fits, and on different conventions on the meaning of the error bars, there is no reason to strictly assume that the ``best model'' is the one with $\chi^2_\red$ closer to 1. Nevertheless, the $\chi^2_\red$ statistics provides a quick way to compare with other works, and it naturally yields numbers of the order of unity. Also, contrary to the values of $\chi^2$, a penalisation for models with too much free parameters is already part of the $\chi^2_\red$ definition. It is a weak penalisation in the context of the data under analysis here, but it is simple. Table \ref{tab:freep} lists the number of free parameters that are used in the $\chi_\red^2$ computation.

\begin{table}
\caption{The number of fitted parameters ($n$) of each model for either a galaxy without or with a bulge. The latter case always has an additional free parameter that corresponds to $\Upsilon_{*B}$. The  values of $n$ listed below only have impact on the computation of $\chi^2_\red$. All the models are subject to the constraints $\Upsilon_{*D}, \Upsilon_{*B} >0$, apart from these, some models have additional constraints.} 
\begin{tabular}{llccc}
\hline
\text{Model} & & $n$  & $n$ & Constrained\\
             & & disk only  & with bulge & parameter \\
\hline
{MOND}           & & 1 & 2 & ---\\
{MOND$_\delta$}  & & 2 & 3 & $ 0.8 \leq \delta \leq 1.2$\\[0.15cm]
{NFW}            & & 3 & 4 & ---\\
{NFW$_{13}$}     & & 3 & 4 & $M_{200} \leq 10^{13} M_\odot$\\
{NFW$_{12}$}     & & 3 & 4 & $M_{200} \leq 10^{12} M_\odot$\\[0.15cm]
{RGGR}           & & 2 & 3 & ---\\
{RGGR$_\delta$}  & & 3 & 4 & $ 0.8 \leq \delta \leq 1.2$\\
\hline
\end{tabular}
\label{tab:freep}
\end{table}

\subsection{On the use of medians} \label{sec:medians}

A significative part of this work consists of deriving rotation curve fits for individual galaxies, the resulting parameters can be found in Table \ref{tab:results}. Nevertheless, it is hard to infer conclusions directly from that table of data. Sometimes a single average quantity is helpful to summarise the behaviour of a certain parameter in the whole sample of data or in a subsample.

Data related to galaxies commonly, for diverse reasons, have large scatter. In particular, consider the quantity $\chi^2_{RD/2}$  for the Sample A. This quantity is present in Table \ref{tab:medianssample} and was derived by a straightforward sum of the values of $\chi^2_{RD/2}$ for all the galaxies that belongs to the Sample A. Clearly the value of this quantity for the RGGR models is significantly lower than the correspond quantities of the other models, but it is impossible to state that this result is representative of RGGR in this sample simply by looking to that number, since a single galaxy dominates this value. For the RGGR and NFW models that quantity reads respectively 605 and 691. Among the 18 galaxies of this sample, there is a single galaxy, NGC 3521, whose individual contribution to $\chi^2_{RD/2}$ reads 411 and 464 respectively for RGGR and NFW. Clearly a simple sum or mean of the individual values of $\chi^2_{RD/2}$ would not lead to a sample representative value. 

In the case of the median, by definition, half of the sample has values lower than it, and the other half has values higher than it. Hence, it is less prone to the influence of outliers and it has a simple expression that better represent the sample results. For similar reasons associated to other quantities, we preferentially compare medians instead mean values.

\section{Results on individual galaxies and analyses} \label{sec:results}

The derived fit parameters for each galaxy and model are in the Table \ref{tab:results}. This table (together with Table \ref{tab:inputdata}) contains all the data necessary to do all the subsequent analyses present in this work. In the Appendix \ref{app:individualgalaxies} the plots for each galaxy considering the models MOND, NFW and RGGR are shown.

\begin{figure*}
\begin{center}
 \includegraphics[width=145mm]{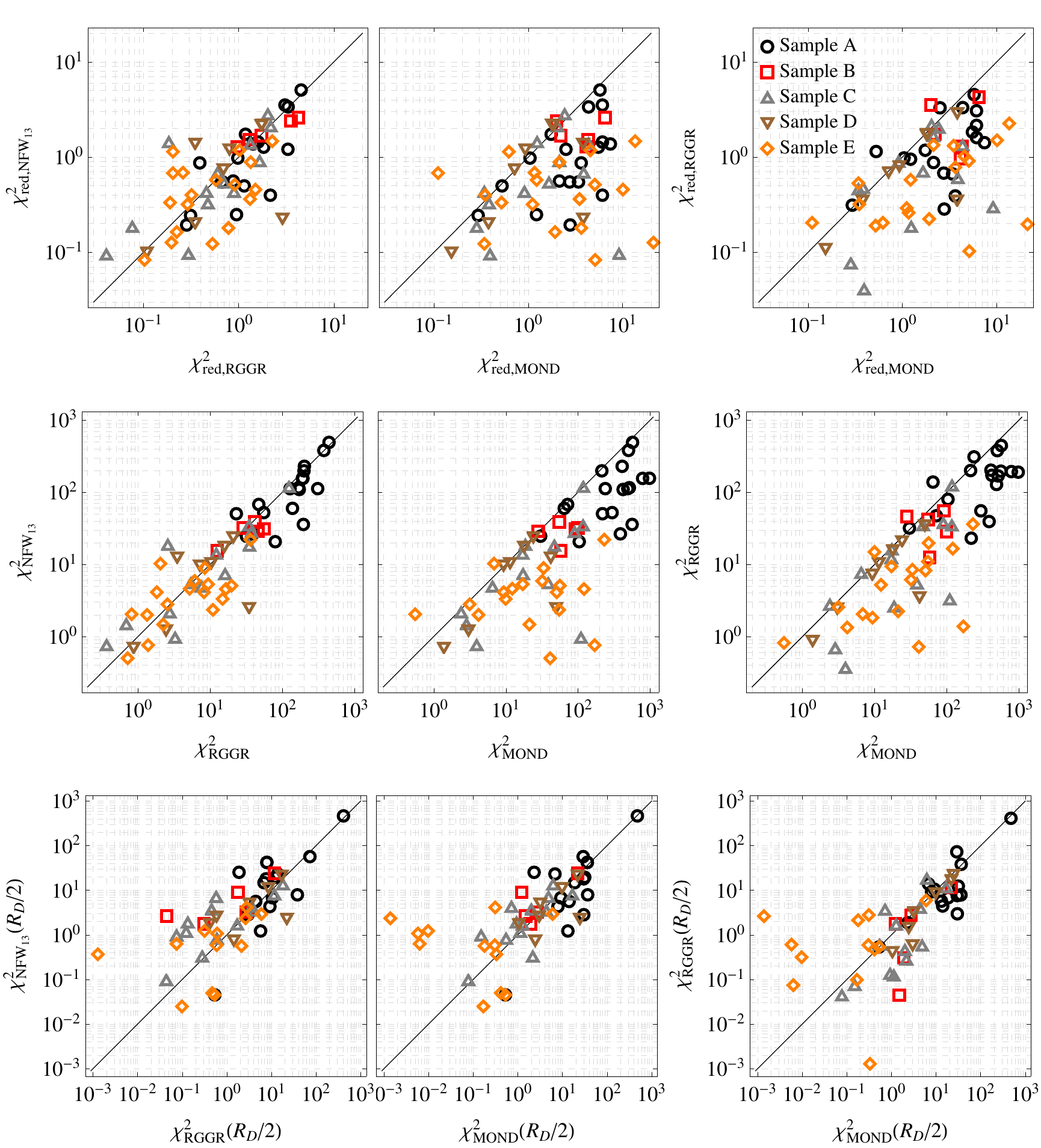}
\end{center}
\caption{Comparison of the values of $\chi^2_\red, \; \chi^2, \; \chi^2_{R_D/2}$ and disk $\Upsilon_*$  for all the galaxies considered and for the models MOND, NFW$_{13}$ and RGGR.  Black circles correspond to Sample A, red squares to Sample B, grey up triangles to Sample C, brown down triangles to Sample D and orange diamonds to Sample E.}
\label{fig:grid2on2comparisonNoDelta}
\end{figure*}

\begin{figure*}
\begin{center}
 \includegraphics[width=145mm]{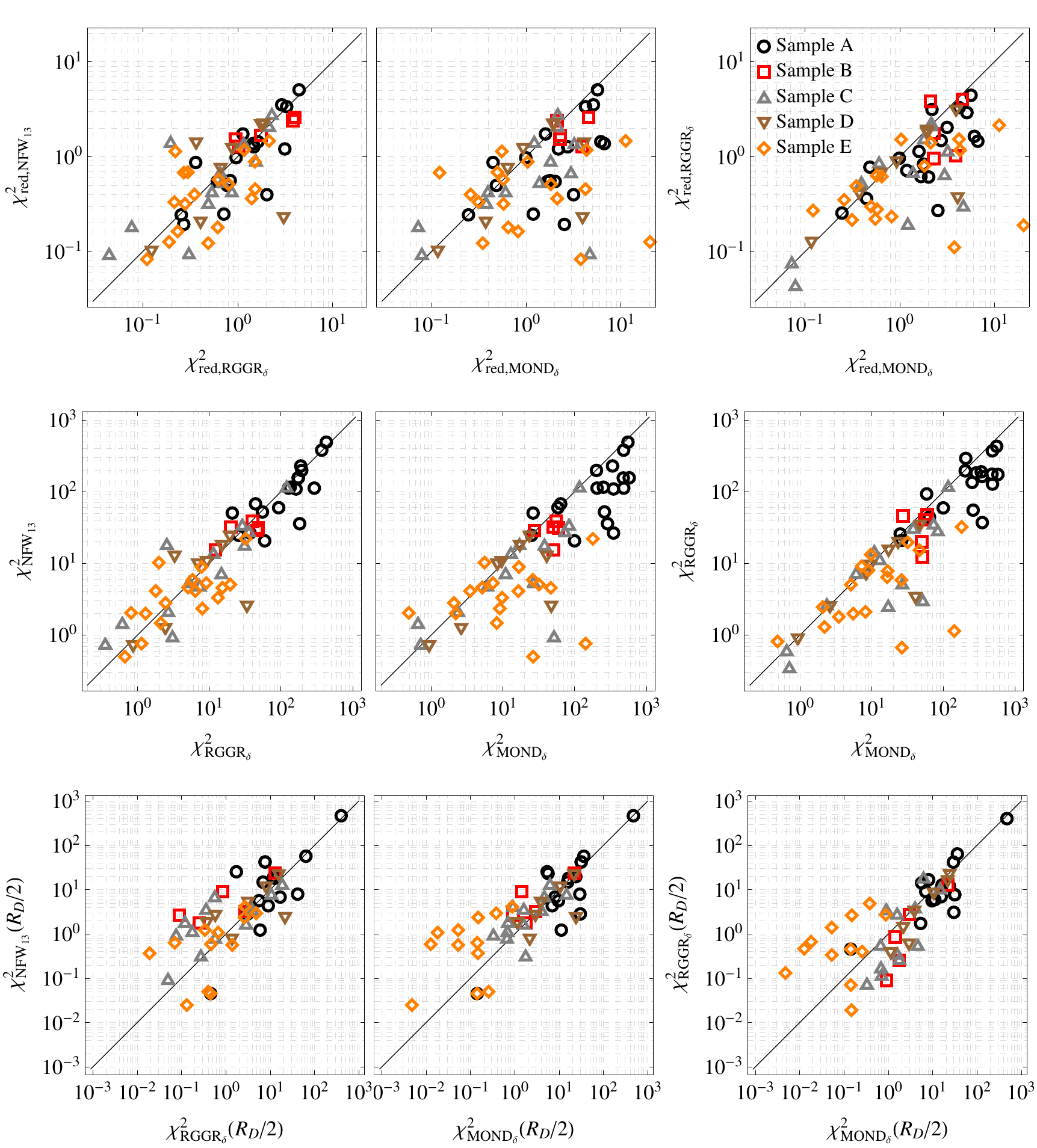}
\end{center}
\caption{The same of Fig. \ref{fig:grid2on2comparisonNoDelta}, but using the models MOND$_\delta$ and RGGR$_\delta$ in place of MOND and RGGR.}
\label{fig:grid2on2comparisonDelta}
\end{figure*}

\begin{figure*}
    \includegraphics[width=\textwidth]{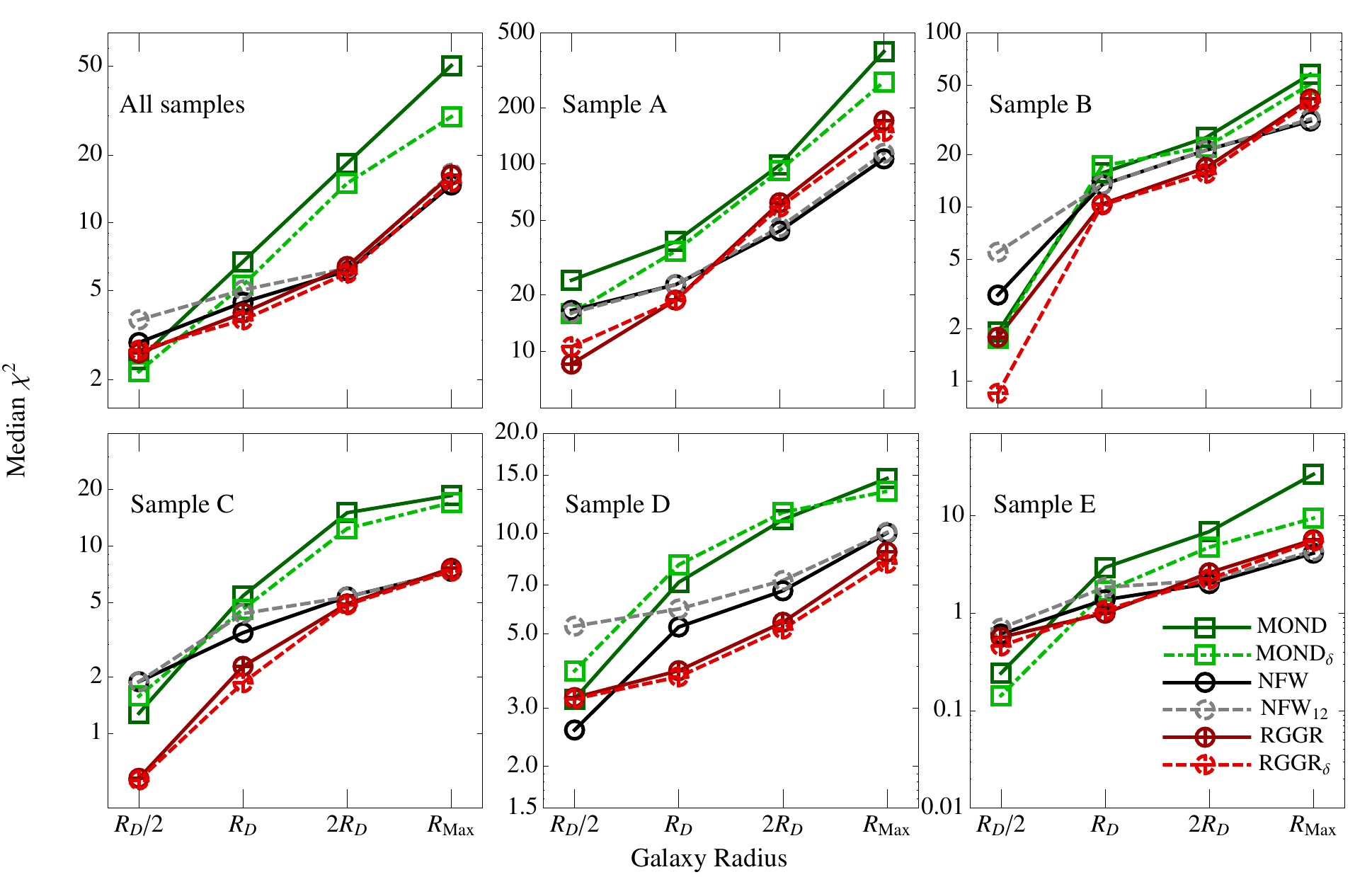}
\caption{Graphical analysis of the medians $\widetilde{\chi^2_{R}}$ for $R = R_D/2, R_D, 2 R_D, R_\mx$ and for the models: MOND (squares, solid dark green lines), MOND$_\delta$ (squares, dot-dashed light green lines), NFW (circles, solid black lines), NFW$_{12}$ (circles, dashed gray lines), RGGR (circles with cross, solid dark red lines) and RGGR$_\delta$ (circles with cross, dashed light red lines). The data shown in these plots is also present in the Tables \ref{tab:medianssample}, \ref{tab:mediansall}.}
    \label{fig:ChiEvol}
\end{figure*}

\begin{figure*}
    \includegraphics[width=\textwidth]{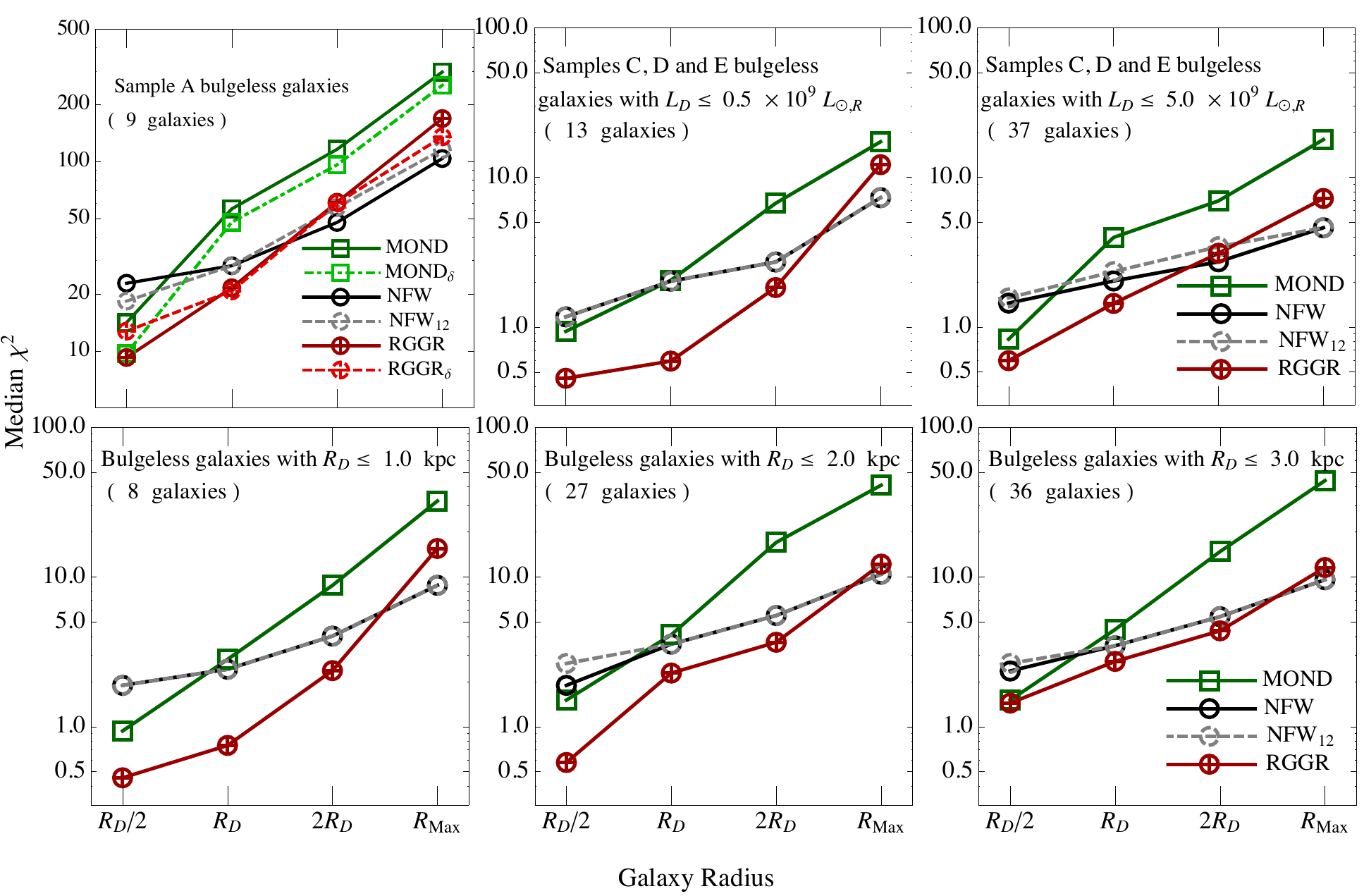}
\caption{Same analysis of Fig. \ref{fig:ChiEvol}, but different samples are considered, with emphasis given to the bulgeless, smaller and less luminous galaxies. The first plot simply considers all the galaxies from the Sample A that have no significant bulge. The second and third plots in the first row considers all the bulgeless galaxies from samples C,D and E with a constraint in $L_D$ in the R band as written in these plots (Samples A and B are not considered since they use a different band for $L_D$). The second row considers all the bulgeless galaxies from all the samples, but with a constrain on $R_D$ as written in the plots. The number of galaxies that satisfy the constraints associated with each of the plots is written in them. Results for MOND$_\delta$ and RGGR$_\delta$ are not shown in the plots with constraints in $L_D$ or $R_D$ since for $\delta \not=1$ these models change the values of $R_D$ and $L_D$. In all these plots, the models NFW and NFW$_{12}$ have the highest values for $\widetilde{\chi^2_{R_D/2}}$.}
    \label{fig:ChiEvolBulgeless}
\end{figure*}

\begin{figure*}
\begin{center}
 \includegraphics[width=175mm]{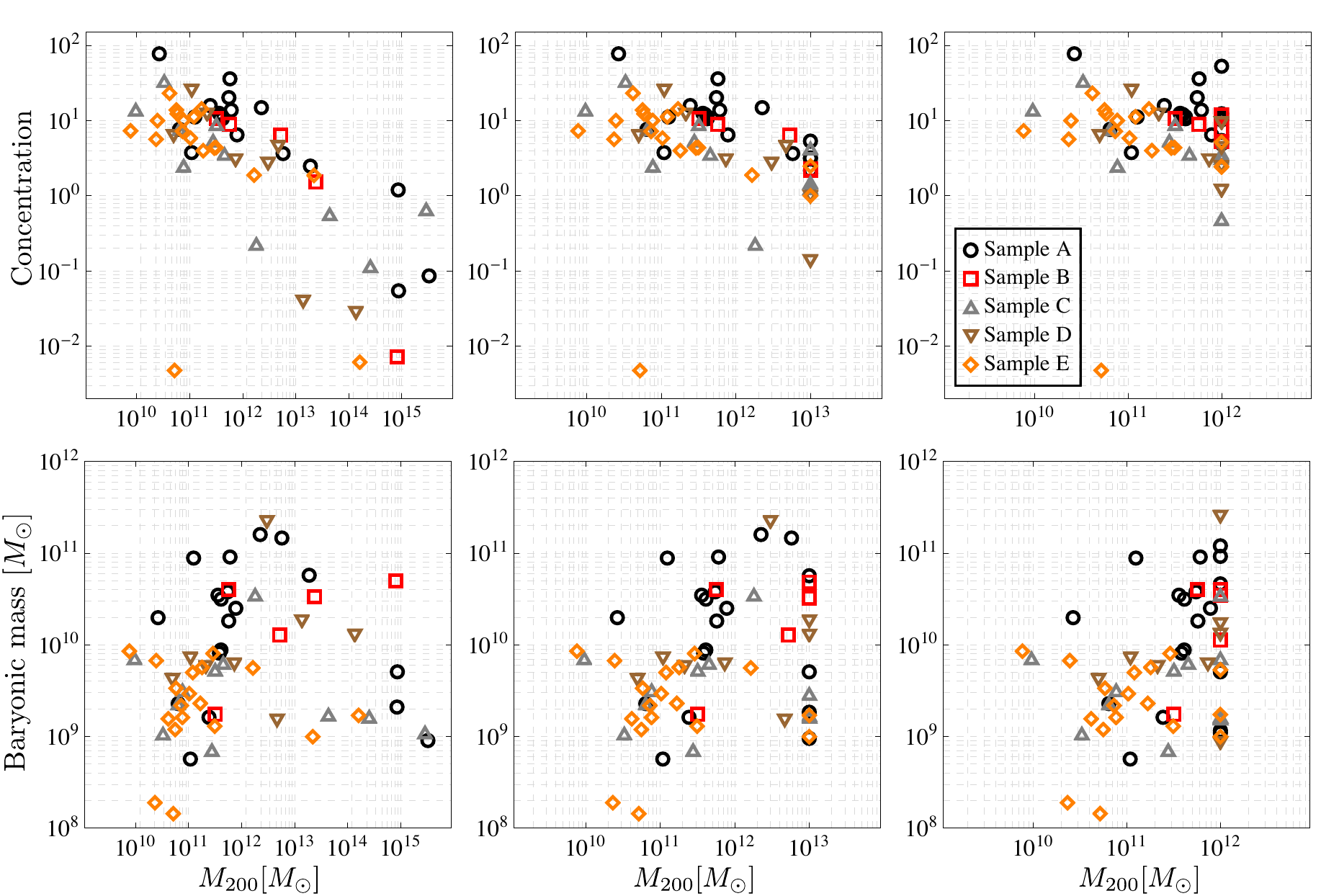}
\end{center}
\caption{Correlation between $M_{200}$ and either the concentration ($c$) or the total baryonic mass of each galaxy. The first, second and third columns correspond respectively to the models NFW, NFW$_{13}$ and NFW$_{12}$. The symbols are the same of Fig.\ref{fig:grid2on2comparisonNoDelta}.}
\label{fig:M200grid}
\end{figure*}

Analyses of the Table \ref{tab:results} results are present in the Tables \ref{tab:medianssample}, \ref{tab:mediansall},  \ref{tab:schifractions}, and in the Figs. \ref{fig:grid2on2comparisonNoDelta}, \ref{fig:grid2on2comparisonDelta}, \ref{fig:ChiEvol}, \ref{fig:ChiEvolBulgeless}, \ref{fig:M200grid}, \ref{fig:barnuBM}.

\begin{table*}
\caption{(The full table is available online). Best fit results for the 62 galaxies and for each of the seven models tested. Explicit results for the models NFW$_{13}$ and NFW$_{12}$ only appear when the fitted $M_{200}$ respectively exceeds $10^{13} M_\odot$ or $10^{12} M_\odot$. Col. (1): the sample. Col.(4)  shows the minimum $\chi^2$ found for each of the fits (all the fits are done such that $\chi^2$ is minimised). Col. (5): the reduced $\chi^2$, see Sec \ref{sec:freep}. Cols. (6)-(8): $\chi^2_{\mbox{\tiny X}}$ stands for $\chi^2$ but considering only the observational data within the galaxy radius $R \leq X$, see Sec.\ref{sec:chi2R}. Cols.(9)-(10) show the disk and bulge stellar mass-to-light ratios in the appropriate band for each sample. Cols. (11)-(12): P1 and P2 stand for model parameters. For NFW and variations they correspond respectively to $r_s$ (kpc) and $\rho_s$ ($M_\odot$ / kpc$^3$), while for RGGR and RGGR$_\delta$ only P1 assumes values and it corresponds to the dimensionless constant $\bar \nu \times 10^{7}$. Col.(13): $\delta$ is the factor that changes the galaxy distance in the models MOND$_\delta$ and RGGR$_\delta$, see Secs. \ref{sec:variations}, \ref{sec:freep}.} 
\begin{tabular}{lllrrrrrrrrrr}
\hline
S & Galaxy   & Model      & $\chi^2$   & $\chi_\red^2$ & $\chi^2_{2R_D}$ & $\chi^2_{R_D}$ & $\chi^2_{R_D/2}$ & $\Upsilon_{*D}$ & $\Upsilon_{*B}$ & \multicolumn{1}{c}{P1} & \multicolumn{1}{c}{P2} & \multicolumn{1}{c}{$\delta$} \\
{(1)} &    {(2)}   &  \multicolumn{1}{c}{(3)}    &   \multicolumn{1}{r}{ (4)}  &  \multicolumn{1}{r}{ (5)}   &  \multicolumn{1}{r}{(6)}   & \multicolumn{1}{r}{(7)}    &    \multicolumn{1}{r}{ (8)}      &  \multicolumn{1}{r}{(9)} &  \multicolumn{1}{c}{(10)} & \multicolumn{1}{r}{(11)} &  \multicolumn{1}{c}{(12)} &  \multicolumn{1}{r}{(13)} \\
\hline
    A       & DDO 154     & MOND       & 218.19 & 3.64   & 13.06   & 3.57    & 2.32     & 0.00      & \multicolumn{1}{c}{---}          & \multicolumn{1}{c}{---}          & \multicolumn{1}{c}{---}          & \multicolumn{1}{c}{---}     \\
    A       & DDO 154     & MOND$_\delta$& 26.35  & 0.45   & 7.95    & 5.57    & 5.37     & 0.58      & \multicolumn{1}{c}{---}          & \multicolumn{1}{c}{---}          & \multicolumn{1}{c}{---}          & 0.80  \\
    A       & DDO 154     & NFW        & 50.42  & 0.87   & 29.53   & 28.14   & 25.27    & 1.25      & \multicolumn{1}{c}{---}          & 27.00      & 5.73$\times 10^5$    & \multicolumn{1}{c}{---}     \\
    A       & DDO 154     & RGGR       & 23.06  & 0.39   & 3.66    & 3.12    & 1.88     & 2.12      & \multicolumn{1}{c}{---}          & 0.20       & \multicolumn{1}{c}{---}          & \multicolumn{1}{c}{---}     \\
    A       & DDO 154     & RGGR$_\delta$ & 21.00  & 0.36   & 2.97    & 2.73    & 1.73     & 1.71      & \multicolumn{1}{c}{---}          & 0.18       & \multicolumn{1}{c}{---}          & 1.20  \\[0.1cm]
    A       & NGC 2403 1D & MOND       & 977.45 & 3.41   & 442.74  & 132.07  & 29.86    & 0.91      & \multicolumn{1}{c}{---}          & \multicolumn{1}{c}{---}          & \multicolumn{1}{c}{---}          & \multicolumn{1}{c}{---}     \\
    A       & NGC 2403 1D & MOND$_\delta$ & 577.88 & 2.02   & 264.08  & 74.21   & 16.33    & 0.58      & \multicolumn{1}{c}{---}          & \multicolumn{1}{c}{---}          & \multicolumn{1}{c}{---}          & 1.20  \\
    A       & NGC 2403 1D & NFW        & 156.24 & 0.55   & 43.16   & 22.19   & 18.05    & 0.39      & \multicolumn{1}{c}{---}          & 14.02      & 7.49$\times 10^6$    & \multicolumn{1}{c}{---}     \\
    A       & NGC 2403 1D & RGGR       & 190.05 & 0.66   & 58.38   & 13.67   & 7.59     & 0.33      & \multicolumn{1}{c}{---}          & 1.73       & \multicolumn{1}{c}{---}          & \multicolumn{1}{c}{---}     \\
    A       & NGC 2403 1D & RGGR$_\delta$ & 174.11 & 0.61   & 60.19   & 18.39   & 12.71    & 0.41      & \multicolumn{1}{c}{---}          & 1.81       & \multicolumn{1}{c}{---}          & 0.80  \\
    \hline
    \end{tabular}
    \label{tab:results}
\end{table*}

\begin{table*}
\caption{Medians and totals for each sample and each model.  The median of a quantity $X$ is denoted by $\widetilde X$. Col. (1): the sample. Col. (3): the median of the reduced $\chi^2$.  Cols. (4)-(7): $\chi^2_{\mbox{\tiny X}}$ stands for the total $\chi^2$ of the sample considering  the observational data within the galaxy radius $R \leq X$, see Sec. \ref{sec:chi2R}.}
\begin{tabular}{clrrrrrrrrrrr}
\hline\\
S & \text{Model} & $\widetilde{\chi^2_\red}$ &  $\chi^2_{R_\mx}$ & $\chi^2_{2R_D}$ & $\chi^2_{R_D}$ & $\chi^2_{R_D/2}$ & $\widetilde{\chi^2_{R_\mx}}$ & $\widetilde{\chi^2_{2R_D}}$ & $\widetilde{\chi^2_{R_D}}$ & $\widetilde{\chi^2_{R_D/2}}$ \\
(1) & (2) & (3) & (4) & (5) & (6) & (7) &(8) & (9) & (10) & (11)  \\
\hline
 A & {MOND} & 3.1 &  6851.2 & 2476.9 & 1203.7 & 744.3 & 397.6 & 98.7 & 38.5 & 24.0 \\
 {A} & {MOND$_\delta$} & 2.1 &  5095.7 & 2025.6 & 1065.4 & 704.1 & 274.1 & 92.5 & 34.3 & 16.0 \\
 {A} & {NFW} & 0.9 &  2385.3 & 1258.7 & 841.7 & 691.2 & 106.4 & 43.9 & 22.7 & 16.5 \\
 {A} & {NFW$_{13}$} & 0.9 &  2398.1 & 1272.6 & 849.7 & 692.4 & 110.8 & 45.9 & 22.7 & 16.4 \\
 {A} & {NFW$_{12}$} & 1.1  & 2996.1 & 1335.1 & 878.4 & 689.9 & 114.0 & 46.0 & 22.7 & 16.0 \\
 {A} & {RGGR} & 1.2  & 2997.4 & 1499.5 & 853.0 & 604.6 & 170.2 & 62.1 & 18.8 & 8.6 \\
 {A} & {RGGR$_\delta$} & 1.1 &  2774.2 & 1458.8 & 832.7 & 600.8 & 148.9 & 58.9 & 18.8 & 10.6 \\[.05in]
 {B} & {MOND} & 4.1 &  331.3 & 151.0 & 87.6 & 29.4 & 57.8 & 25.1 & 15.7 & 1.9 \\
 {B} & {MOND$_\delta$} & 2.3 &  243.0 & 123.1 & 76.5 & 29.2 & 50.8 & 22.1 & 17.2 & 1.8 \\
 {B} & {NFW} & 1.6 &  143.4 & 91.7 & 67.8 & 39.7 & 31.1 & 21.2 & 13.4 & 3.1 \\
 {B} & {NFW$_{13}$} & 1.7 &  145.8 & 93.4 & 69.8 & 40.5 & 31.1 & 21.2 & 13.4 & 3.2 \\
 {B} & {NFW$_{12}$} & 2.6 &  183.7 & 117.2 & 94.6 & 53.1 & 32.0 & 21.2 & 13.4 & 5.5 \\
 {B} & {RGGR} & 1.7 & 184.1 & 89.5 & 49.1 & 16.4 & 41.8 & 16.8 & 10.4 & 1.8 \\
 {B} & {RGGR$_\delta$} & 1.8 &  166.9 & 81.9 & 45.7 & 16.8 & 40.3 & 15.6 & 10.3 & 0.8 \\[.05in]
 {C} & {MOND} & 1.7 & 589.1 & 309.7 & 183.1 & 41.3 & 18.6 & 15.0 & 5.4 & 1.3 \\
 {C} & {MOND$_\delta$} & 1.4  & 440.7 & 261.9 & 162.6 & 37.9 & 17.0 & 12.4 & 4.6 & 1.6 \\
 {C} & {NFW} & 0.5 &  252.6 & 180.1 & 118.8 & 47.1 & 7.3 & 5.3 & 3.4 & 1.9 \\
 {C} & {NFW$_{13}$} & 0.5 &  255.0 & 181.7 & 120.0 & 47.1 & 7.3 & 5.3 & 3.7 & 1.9 \\
 {C} & {NFW$_{12}$} & 0.5 &  262.5 & 186.8 & 124.9 & 48.9 & 7.3 & 5.3 & 4.4 & 1.9 \\
 {C} & {RGGR} & 0.6 &  281.6 & 201.6 & 121.0 & 41.3 & 7.6 & 4.9 & 2.3 & 0.6 \\
 {C} & {RGGR$_\delta$} & 0.7 &  264.4 & 191.6 & 115.0 & 41.9 & 7.4 & 4.9 & 1.9 & 0.6 \\[.05in]
 {D} & {MOND} & 1.4 &  158.4 & 112.7 & 91.0 & 66.0 & 14.6 & 11.0 & 7.1 & 3.2 \\
 {D} & {MOND$_\delta$} & 1.4 &  150.2 & 112.6 & 91.7 & 66.5 & 13.4 & 11.6 & 8.1 & 3.9 \\
 {D} & {NFW} & 1.0 &  78.0 & 66.0 & 53.2 & 44.5 & 10.0 & 6.7 & 5.2 & 2.6 \\
 {D} & {NFW$_{13}$} & 1.0  & 78.1 & 66.2 & 53.4 & 44.7 & 10.0 & 6.7 & 5.2 & 2.6 \\
 {D} & {NFW$_{12}$} & 1.0  & 86.1 & 74.4 & 60.3 & 50.7 & 10.1 & 7.2 & 5.9 & 5.3 \\
 {D} & {RGGR} & 0.7 & 95.7 & 81.2 & 64.6 & 53.2 & 8.8 & 5.4 & 3.9 & 3.2 \\
 {D} & {RGGR$_\delta$} & 0.8 & 90.9 & 77.2 & 61.6 & 50.8 & 8.1 & 5.2 & 3.7 & 3.2 \\[.05in]
 {E} & {MOND} & 2.0 &  873.0 & 315.8 & 124.3 & 8.3 & 26.5 & 6.8 & 2.9 & 0.2 \\
 {E} & {MOND$_\delta$} & 0.7  & 538.9 & 225.2 & 88.6 & 2.2 & 9.4 & 4.7 & 1.7 & 0.1 \\
 {E} & {NFW} & 0.4 &  89.8 & 48.0 & 33.0 & 13.9 & 4.1 & 2.0 & 1.4 & 0.6 \\
 {E} & {NFW$_{13}$} & 0.4 & 90.0 & 48.2 & 33.2 & 14.0 & 4.1 & 2.0 & 1.4 & 0.6 \\
 {E} & {NFW$_{12}$} & 0.5  & 92.2 & 50.2 & 34.8 & 14.6 & 4.3 & 2.2 & 1.8 & 0.7 \\
 {E} & {RGGR} & 0.4 & 148.4 & 58.0 & 37.3 & 16.2 & 5.7 & 2.6 & 1.0 & 0.6 \\
 {E} & {RGGR$_\delta$} & 0.4  & 134.9 & 52.7 & 33.8 & 14.1 & 5.5 & 2.2 & 1.1 & 0.5 \\
 \hline
\end{tabular}
\label{tab:medianssample}
\end{table*}

\begin{table*}
\caption{Medians and totals for each model considering all the 62 galaxies. Same symbols of Table \ref{tab:medianssample}.}
\begin{tabular}{lrrrrrrrrrrr}
\hline\\
\text{Model} & $\widetilde{\chi^2_\red}$ & $\chi^2_{R_\mx}$ & $\chi^2_{2R_D}$ & $\chi^2_{R_D}$ & $\chi^2_{R_D/2}$ & $\widetilde{\chi^2_{R_\mx}}$ & $\widetilde{\chi^2_{2R_D}}$ & $\widetilde{\chi^2_{R_D}}$ & $\widetilde{\chi^2_{R_D/2}}$ \\
(1) & (2) & (3) & (4) & (5) & (6) & (7) & (8) &(9) & (10)  \\
\hline
{MOND} & 2.1  & 8803.0 & 3366.2 & 1689.7 & 889.3 & 50.1 & 18.4 & 6.7 & 2.5 \\
{MOND$_\delta$} & 1.8  & 6468.5 & 2748.5 & 1484.7 & 840.0 & 29.7 & 15.0 & 5.3 & 2.2 \\
{NFW} & 0.7  & 2949.1 & 1644.5 & 1114.5 & 836.5 & 14.7 & 6.1 & 4.4 & 2.9 \\
{NFW$_{13}$} & 0.7 &  2967.1 & 1662.0 & 1125.9 & 838.7 & 14.7 & 6.1 & 4.4 & 2.9 \\
{NFW$_{12}$} & 0.7 &  3620.6 & 1763.7 & 1193.0 & 857.3 & 16.6 & 6.3 & 5.0 & 3.7 \\
{RGGR} & 0.9 & 3707.3 & 1929.8 & 1125.1 & 731.7 & 16.3 & 6.4 & 4.0 & 2.6 \\
{RGGR$_\delta$} & 0.8  & 3431.4 & 1862.2 & 1088.8 & 724.6 & 15.2 & 6.0 & 3.7 & 2.7 \\
\hline
\end{tabular}
\label{tab:mediansall}
\end{table*}

\subsection{$\chi^2$ and $\chi^2_\red$ results} \label{sec:chiresults}

A graphical comparison of the minimum $\chi^2$ values of each galaxy can be find in the second line of Figs. \ref{fig:grid2on2comparisonNoDelta}, \ref{fig:grid2on2comparisonDelta}. Both the  total $\chi^2$ per sample (i.e., the sum of the minimum $\chi^2$ of each galaxy for a given sample)  and their medians per sample are shown in Table \ref{tab:medianssample}. The total and median of $\chi^2$ for all the galaxies is in Table \ref{tab:mediansall}, and the latter median is also depicted in Fig. \ref{fig:ChiEvol}.

From the above, it should be clear that there is a general trend for the $\chi^2$ values of the models MOND, NFW and RGGR, namely, $\chi^2_{\mbox{\tiny NFW}} \lesssim \chi^2_{\mbox{\tiny RGGR}}< \chi^2_{\mbox{\tiny MOND}}.$ That is, with a few galaxy exceptions, the minimum $\chi^2$ values that can be achieved by RGGR are clearly smaller than those of MOND; whilst the $\chi^2$ values of NFW and RGGR are rather close, but both the total and the medians slightly favours the NFW model. The above trend does not change significantly if the four variations on these three models are used. It should be stressed that these models have different number of free parameters, see Table \ref{tab:freep}.

Among all the variations, the one with the highest impact on $\chi^2$ is MOND$_\delta$, which significantly reduces the value of $\chi^2$ when compared with MOND. 

Even the stronger constrained version of the NFW model, NFW$_{12}$, lead to no significant changes on the values of $\chi^2$. The highest impact this constrain lead to was on the rotation curve close to the galaxy centre, to be discussed in Sec. \ref{sec:chiRresults}. Considering the values of $\chi^2$, the sample that most felt the difference between NFW and NFW$_{12}$ was Sample A. This sample has five galaxies with baryonic masses about $10^{11} M_\odot$, hence it is natural that this sample was especially affected by that constraint.

The reduced chi-squared $\chi^2_\red$ has a small compensation for the number of fitted parameters, and it changes only slightly the picture above described for $\chi^2$. We only add that since the values of $\chi^2_{\mbox{\tiny RGGR$_\delta$}}$ are only slightly lower than those of $\chi^2_{\mbox{\tiny RGGR}}$, for diverse cases the values of $\chi^2_{\mbox{\tiny red, RGGR$_\delta$}}$ are slightly higher than those of $\chi^2_{\mbox{\tiny red, RGGR}}$.


\subsection{$\chi^2_R$ results} \label{sec:chiRresults}

Considering the results of all the galaxies, both the total and median of $\chi^2$ is smaller for the NFW model than the corresponding values of RGGR and MOND, which is expected since the last two models have fewer parameters. Notwithstanding,  the relative differences between their values of $\chi^2_R$ dramatically decrease, and at some point change the sign, when one considers the radii $R = R_D$ and, in particular, $R = R_D/2$. The NFW$_{12}$ model do not change significantly the value of $\widetilde{\chi^2}$, but clearly systematically increases the value of $\widetilde{\chi^2_{RD/2}}$, see Figs. \ref{fig:ChiEvol}, \ref{fig:ChiEvolBulgeless}.

A graphical comparison of the $\chi^2_{RD/2}$ values of each galaxy can be find in the third line of Figs. \ref{fig:grid2on2comparisonNoDelta}, \ref{fig:grid2on2comparisonDelta}. Both the  total $\chi^2_{R}$ per sample and their medians per sample are shown in Table \ref{tab:medianssample}. The total and median of $\chi^2_R$ for all the galaxies is in Table \ref{tab:mediansall}, while the medians are also shown in Fig. \ref{fig:ChiEvol}.

Considering the Fig. \ref{fig:ChiEvol},  there is a single sample in which the median (and the total) value of $ \chi^2_{RD/2, \mbox{\tiny NFW}}$ was the lowest among all the models, which is Sample D. This sample has eight galaxies, and two of these galaxies yielded very low concentrations together with very high $M_{200}$ (equivalently, too high $r_s$) (see Fig. \ref{fig:M200grid}). That is, in this sample the NFW model could achieve better concordance for $\chi^2_{RD/2}$, but at the expense of using unreasonable parameters. On applying the constraint $M_{200} \leq 10^{12} M_\odot$, the fitted value of $r_s$ was lowered and the effect of the cusp appeared more clearly (even without the need of imposing a correlation between $c$ and $M_{200}$). 

Figures \ref{fig:grid2on2comparisonNoDelta}, \ref{fig:grid2on2comparisonDelta} also show that there is a tendency for both MOND and RGGR to derive smaller values of $\chi^2_{RD/2}$ than NFW.

\bigskip

Table \ref{tab:schifractions} shows a different and complementary approach of evaluating systematics on the inner radii phenomenology of galaxies. The previous analyses focused on direct comparisons between the values of the ``truncated'' chi-squared values, that is, on $\chi^2_R$, hence the addressed question can be written as: ``which model has the best concordance with the inner radii observational data?''. This other approach deals with another question: ``which model better fits the inner radii region in comparison with the fits of the same model at higher radii?''.

\begin{table}
\caption{Medians of the quantities $\chi^2_{R_D}/ \chi^2_{R_D/2} - 1$ and $\chi^2_{2 R_D}/\chi^2_{R_D/2} - 1$ considering all the galaxies from all the samples such that $\chi^2_{R_D/2} >0$ (i.e., 52 galaxies). If one assumes high density and homogeneous distribution of observational data along all the galaxy rotation curves, one would expect that a model with no bias towards any part of the galaxies would respectively yield $\sim 1.0$ and $\sim 3.0$ for the cited medians. Since in most galaxies there is less observational data inside $R < R_D/2$ than in $ R_D/2 < R < R_D$, actually the previous expectations should be respectively changed to $\gtrsim 1.0$ and $\gtrsim 3.0$.} 
\begin{tabular}{llcc}
\hline\\
\text{Model} & & Med$\(\frac{\chi^2_{R_D}}{\chi^2_{R_D/2}} - 1\)$ & Med$\(\frac{\chi^2_{2 R_D}}{\chi^2_{R_D/2}} - 1\)$ \\[0.5cm]
\hline
{MOND}           & & 1.2 & 6.6 \\
{MOND$_\delta$}  & & 1.3 & 6.9 \\
{NFW}            & & 0.3 & 1.1 \\
{NFW$_{13}$}     & & 0.4 & 1.3 \\
{NFW$_{12}$}     & & 0.4 & 1.2 \\
{RGGR}           & & 0.9 & 3.6 \\
{RGGR$_\delta$}  & & 0.8 & 3.0 \\
\hline
Expected         & &$\gtrsim 1.0$ & $\gtrsim 3.0$\\
\hline
\end{tabular}
\label{tab:schifractions}
\end{table}

If one considers the existence of a galaxy rotation curve with high density and homogeneous distribution of observational data with the same error bars along all the  rotation curves, one would expect that a model with no bias towards any part of the galaxies would yield $\chi^2_{R_D}/\chi^2_{R_D/2} = 2$, or equivalently that
\be
    \frac{\chi^2_{R_D} - \chi^2_{R_D/2}}{\chi^2_{R_D/2}} = 1.
    \label{eq:chifractions}
\ee
For this hypothetical rotation curve, the quantities $\chi^2_{R_D}$ and $\chi^2_{R_D/2}$ are correlated (in particular $\chi^2_{R_D} > \chi^2_{R_D/2}$ always), but $\chi^2_{R_D} - \chi^2_{R_D/2}$ and $\chi^2_{R_D/2}$ are independent quantities {\it a priori}. For a model that systematically fits better the RC data in the radius $R < R_D/2$, in comparison with the data in $ R_D/2< R < R_D$, one would expect that, for the hypothetical rotation curve explained above, $\frac{\chi^2_{R_D} - \chi^2_{R_D/2}}{\chi^2_{R_D/2}} > 1$.

It is easy to check that the RCs studied here systematically deviate from the above hypothetical picture, since typically there are less observational data points inside $R<R_D/2$ than in the region $R_D/2 < R < R_D$. Therefore, considering a sample of galaxies with this bias, for a model with no bias towards any particular galaxy region, it is expected that 
\ba
     \label{eq:chifractionsExpect}
    &&\mbox{Med}\(\frac{\chi^2_{R_D} - \chi^2_{R_D/2}}{\chi^2_{R_D/2}} \) \gtrsim  1,\\[.01in]
    &&\mbox{Med}\(\frac{\chi^2_{2R_D} - \chi^2_{R_D/2}}{\chi^2_{R_D/2}} \) \gtrsim  3.
\ea
where we use Med$(X)$ for the median of $X$. If the above inequalities are not satisfied, the model has a bias towards poorly fitting the region with $R < R_D/2$, in comparison with regions with greater radii. This is the case of NFW, as shown in the Table \ref{tab:schifractions}.


\subsection{Other parameters results} \label{sec:Yresults}

Figure \ref{fig:M200grid} shows the relations between $M_{200}$, $c$ and the total baryonic mass (stellar and gas) of each galaxy, considering the best-fits derived for NFW, NFW$_{13}$ and NFW$_{12}$. A correlation between the derived values of $M_{200}$ and $c$ can be spotted, and, as expected from the simulations results, typically $c$ decreases with the increase of $M_{200}$. The dispersion becomes significantly higher for $c < 1$ or $M_{200} > 10^{13} M_\odot$, which are also values that are unphysical considering the simulations and the galaxies in these samples. The constrained variations NFW$_{13}$ and NFW$_{12}$ do not pose {\it a priori} any correlation between $c$ and $M_{200}$, but the best-fit results are such that favor the correlation, in particular the low $c$ values are almost eliminated.

Regarding the relation between the total baryonic mass and $M_{200}$, no significant correlation can be spotted in Fig. \ref{fig:M200grid}. The second line of the same figure also shows that some galaxies whose derived $M_{200}$ is higher than $10^{13} M_\odot$ are galaxies with baryonic mass about $10^{9} M_\odot$, which leads to very high  discrepancies between dark to luminous matter, from four to six orders of magnitude. 

\bigskip

\begin{figure}
\begin{center}
 \includegraphics[width=60mm]{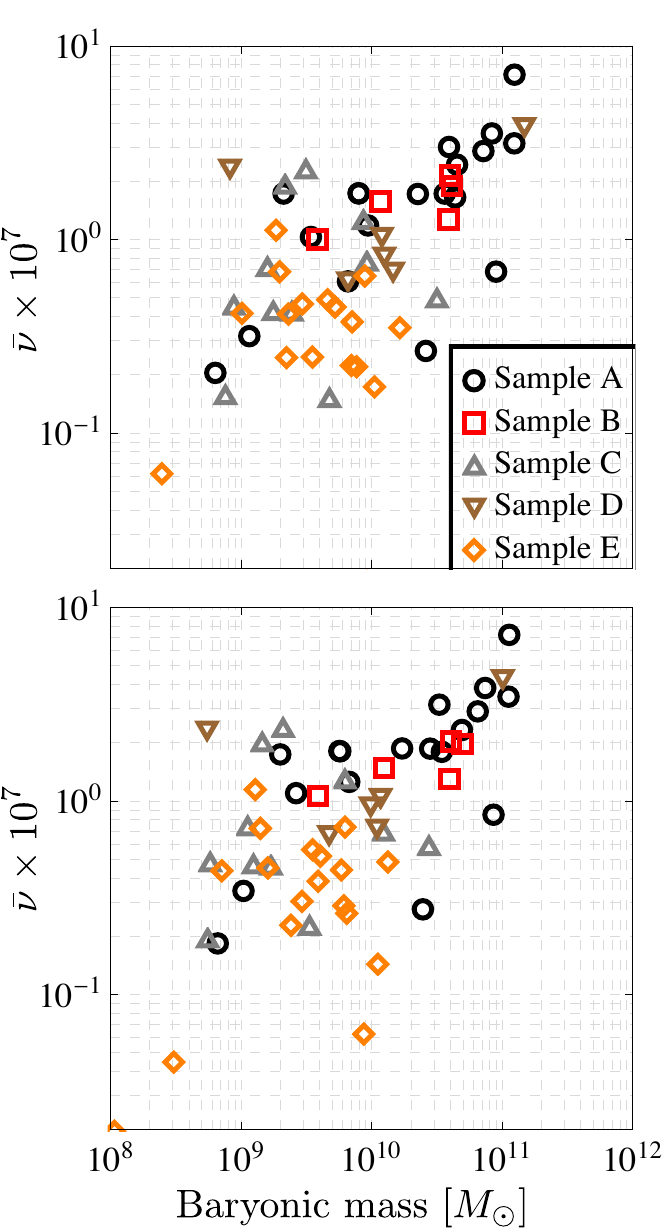}
\end{center}
\caption{Correlation between $\bar \nu$ and the total baryonic mass. The top and bottom plots correspond respectively to RGGR and RGGR$_\delta$.  The symbols are the same of Fig.\ref{fig:grid2on2comparisonNoDelta}. Four galaxies with $\bar \nu = 0$ and two galaxies with $0< \bar \nu < 10^{-9}$ are not shown.}
\label{fig:barnuBM}
\end{figure}

Figure \ref{fig:barnuBM} shows the the existence of a correlation, with large dispersion, between the RGGR dimensionless parameter $\bar \nu$ and the total baryonic mass.  As expected, $\bar \nu$ typically increases as the mass of the system increases. Additional correlations of $\bar \nu$ and baryonic parameters are beyond the scope of this work.

\section{Conclusions}

According to  N-body simulations, it is expected that the density profile for dark matter (DM) is similar to the Navarro-Frenk-White (NFW) profile (eq. \ref{nfwprofile}) \citep{1996ApJ...462..563N, Navarro:1996gj, 0521857937}. One of the key features of this profile is the existence of a density cusp at the galaxy centre ($r=0$). The NFW profile formally depends on two parameters, but there is a correlation between these that depends on cosmology. If this correlation is used in order to eliminate one parameter in favour of the other, then some examples are known whose resulting rotation curve is clearly unsatisfactory \citep[e.g.,][]{Gentile:2004tb, 2005ApJ...634L.145G, Gentile:2006hv}. Nevertheless, to use such correlation as if it were an exact expression raises additional issues, in particular since a significative dispersion is found in the simulations. 

Here we explore the cusp physical consequences without imposing a correlation between the NFW parameters. By a direct inspection of the rotation curve (RC) plots in Fig.\ref{fig:Things1Dplots1} it is not easy to develop comparisons on the RCs trend close to the centre. To this end the quantities $\chi^2_{2RD}, \chi^2_{RD}$ and $\chi^2_{RD/2}$ were introduced in Sec. \ref{sec:chi2R}. A similar approach was used by \cite{deBlok:2002tg}, and they shown that, even without imposing the mentioned correlation, the NFW profile is not favoured when compared to a cored DM profile. 

Instead of comparing the NFW fits to other models with DM halos, we tackled the cusp issue by comparing the resulting NFW rotation curves to the resulting rotation curves of   modified gravity models without DM. These models have no DM density cusp, however they may  have systematic issues on their rotation curves that also pose observational problems at radii close to the galactic centre. Two modified gravity models are here studied and confronted with the NFW results, namely MOND and RGGR (see Sec. \ref{sec:models} for a review on these models). The procedures used here to compare these models can be also applied to other modified gravity models.

Two main issues were addressed in this work:  to introduce a method for evaluating if there is a galactic radius below which a given model can agree with the observational data better than NFW, and  to significantly extend the sample on which RGGR has been tested and compared to other approaches.

By analysing 62 galaxies from five samples, we confirm that the NFW profile has a systematical tendency of generating poor RC fits close to the galaxy centre, when compared with the modified gravity models here studied. This tendency is stronger for ``smaller'' galaxies. In particular, we find here that there is a radius, given by half the disk scale length ($R_D/2$), below which both RGGR and MOND can match the data about as well or better than NFW, albeit the formers have fewer free parameters. This behaviour is in general enhanced when considering a bulgeless subsample, with either the galaxies with smaller disk scale length, or those with lower luminosity (see Figs. \ref{fig:ChiEvol}, \ref{fig:ChiEvolBulgeless}).  Considering the complete rotation curve data, RGGR could achieve fits with better agreement than MOND, and almost as good as a NFW halo with two free parameters (NFW and RGGR have respectively two and one more free parameters than MOND).

Besides the above main results, we have also evaluated four variations on the models above, namely NFW$_{13}$, NFW$_{12}$, RGGR$_\delta$ and MOND$_\delta$ (see Sec. \ref{sec:variations}). In particular we found that a constrain in $M_{200}$ is sufficient for eliminating many of the cases with too low concentrations ($c$), and may have negligible impact on most of the galaxy fits. We confirm that variations on the galaxy distance of the $20\%$ order are sufficient to significantly improve the MOND results, and we find that the improvements for RGGR were rather modest, in many cases negligible. Finally, we directly evaluated for the first time the existence of some correlation between baryonic mass and the $\bar \nu$ parameter of RGGR (see Fig. \ref{fig:barnuBM}).

To conclude, we rephrase one of our conclusions in the following way:  if the rotation curves derived from a baryonic model with a 2-parameter NFW DM halo are considered to be too discrepant with the observations at the centre of galaxies, then, according to these results, this is not a sufficiently strong restriction for dismissing either MOND or RGGR.

\section*{Acknowledgements}
We thank I. L. Shapiro for important discussions on the Renormalization Group and its possible consequences to gravity at galactic scales, and to W.J.G. de Blok and R. Swaters for kindly providing part of the data used in this work. DCR and JCF also thank CNPq (Brazil) and FAPES (Brazil) for partial support. PLO also thanks CAPES (Brazil) for support.

\bibliographystyle{mn2e} 

\bibliography{bibdavi2014B}{}

\appendix

\section{Details on individual galaxies} \label{app:individualgalaxies}

Figure \ref{fig:Things1Dplots1} shows the rotation curves of all the 62 galaxies for the models MOND, NFW and RGGR.

\newcommand{\opPlots}[1]{\includegraphics[trim = 0cm 1.7cm 0cm 0.5cm, clip=true, width=180mm]{#1}}
\newcommand{\opPlotsF}[1]{\includegraphics[trim = 0cm 0.7cm 0cm 0.5cm, clip=true, width=180mm]{#1}}


\begin{figure*}
\begin{subfigure}
  \centering
  \opPlots{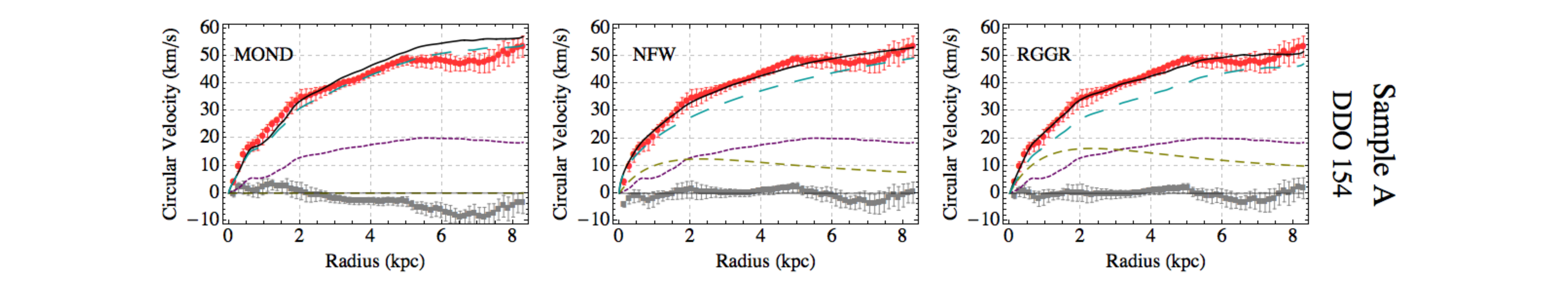}
\end{subfigure}%
\begin{subfigure}
  \centering
  \opPlots{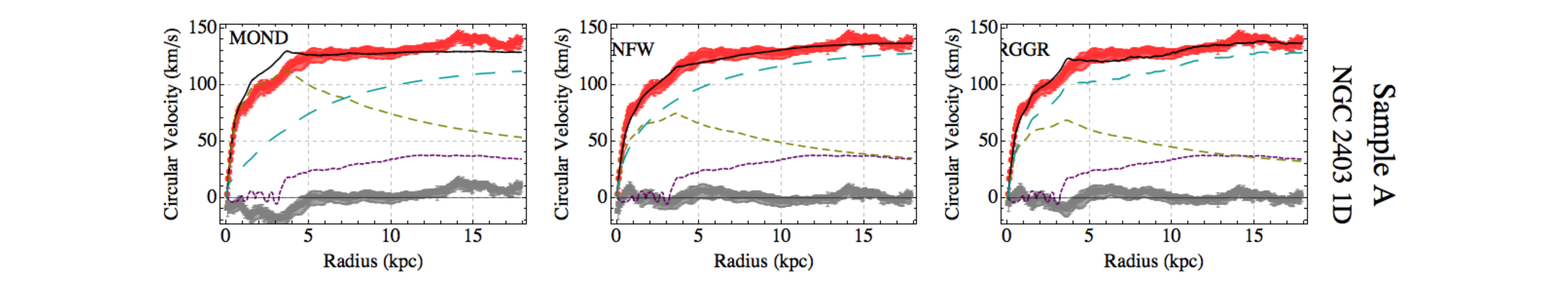}
\end{subfigure}
\begin{subfigure}
  \centering
  \opPlots{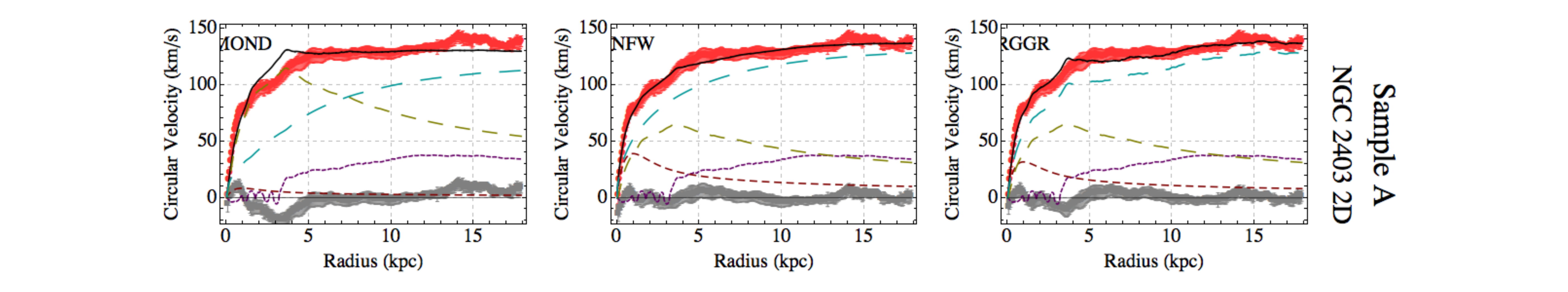}
\end{subfigure}
\begin{subfigure}
  \centering
  \opPlots{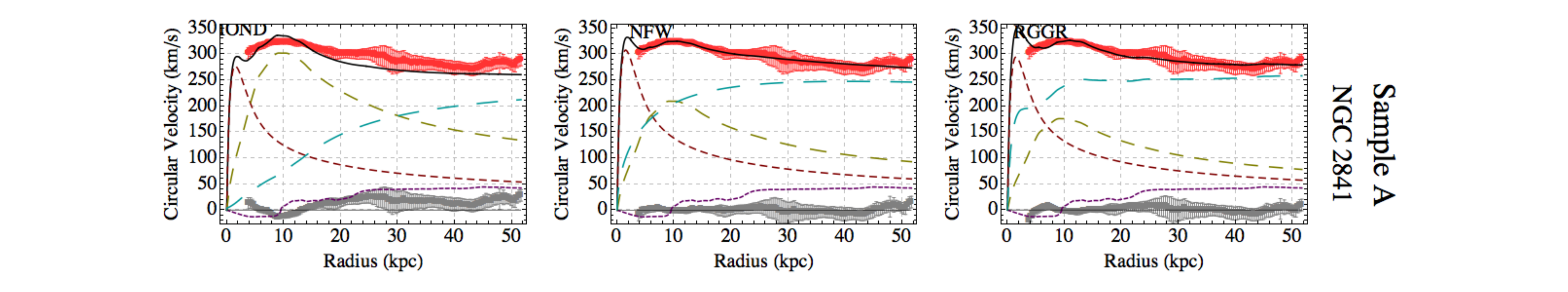}
\end{subfigure}
\begin{subfigure}
  \centering
  \opPlots{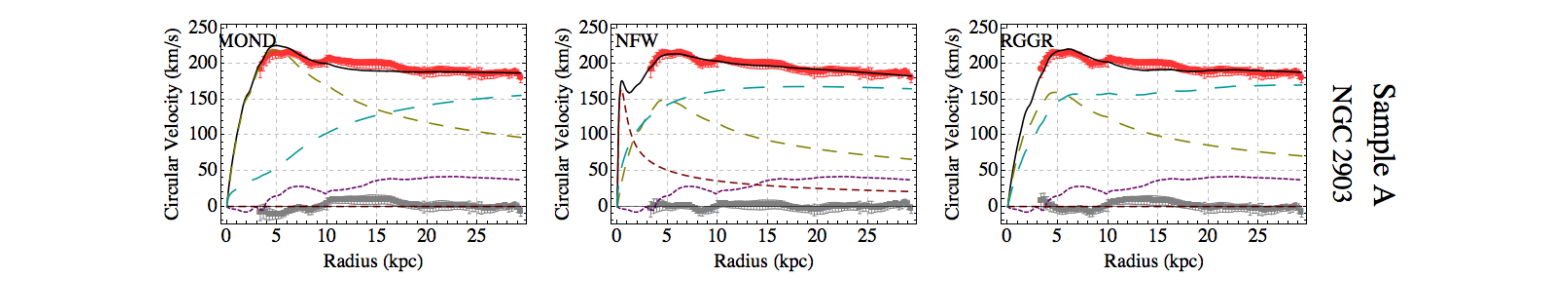}
\end{subfigure}
\begin{subfigure}
  \centering
  \opPlots{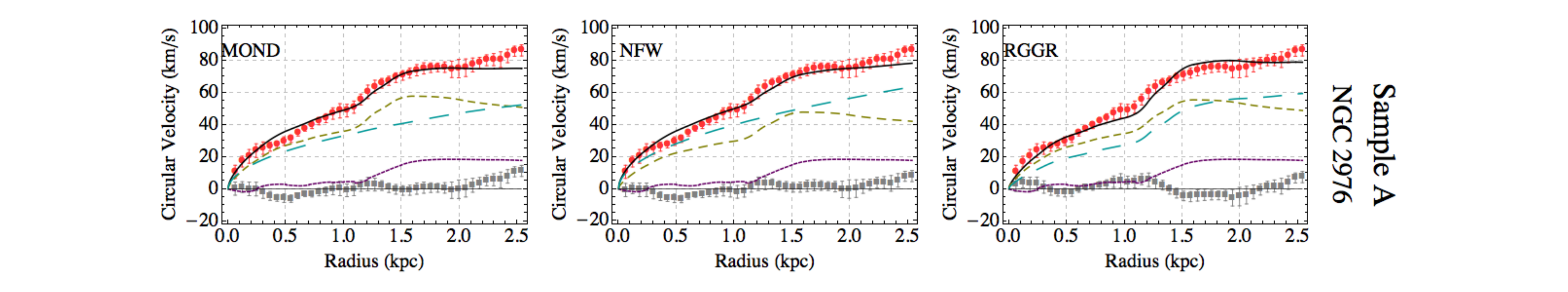}
\end{subfigure}
\begin{subfigure}
  \centering
  \opPlotsF{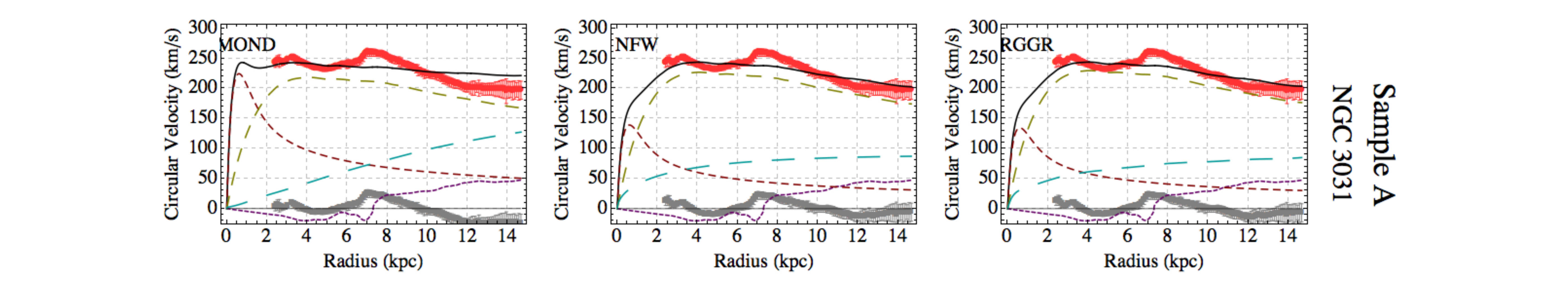}
\end{subfigure}
\caption{(Higher quality images can be found in the on-line material of the published version) Plots show galaxy by galaxy rotation curves (RCs) fits for the models MOND, NFW and RGGR, and for the Samples A, B, C, D and E. The red dots with error bars are the observational RC  data. The squares with error bars are the residues of the fits. The black solid curve is the best fit RC of the model, whose decomposition is present in the following dashed curves:  stellar disk (yellow, dashed), stellar bulge (dark red, dashed), gas (purple, short-dashed) and dark matter for NFW or non-Newtonian contribution for MOND and RGGR (cyan, long-dashed). The graphical appearance of the RCs derived for the models MOND$_\delta$, NFW$_{12}$, NFW$_{13}$ and RGGR$_\delta$ is in most cases very similar to the RCs shown here, hence the plots of the latter models are not displayed.} \label{fig:Things1Dplots1}
\end{figure*}

\begin{figure*}
\begin{subfigure}
  \centering
  \opPlots{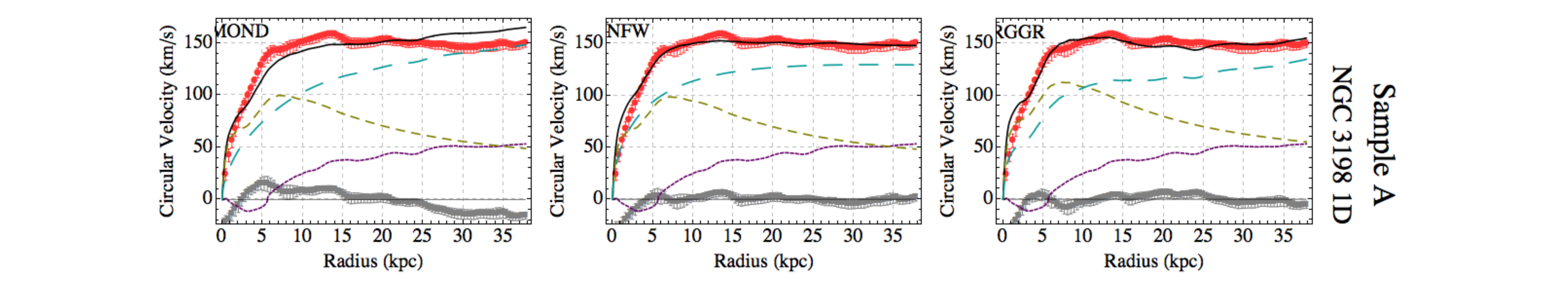}
\end{subfigure}
\begin{subfigure}
  \centering
  \opPlots{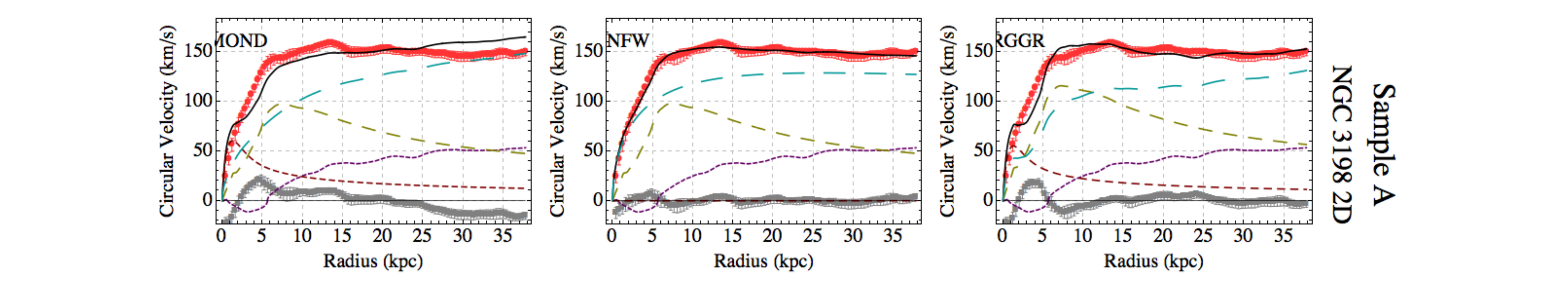}
\end{subfigure}
\begin{subfigure}
  \centering
  \opPlots{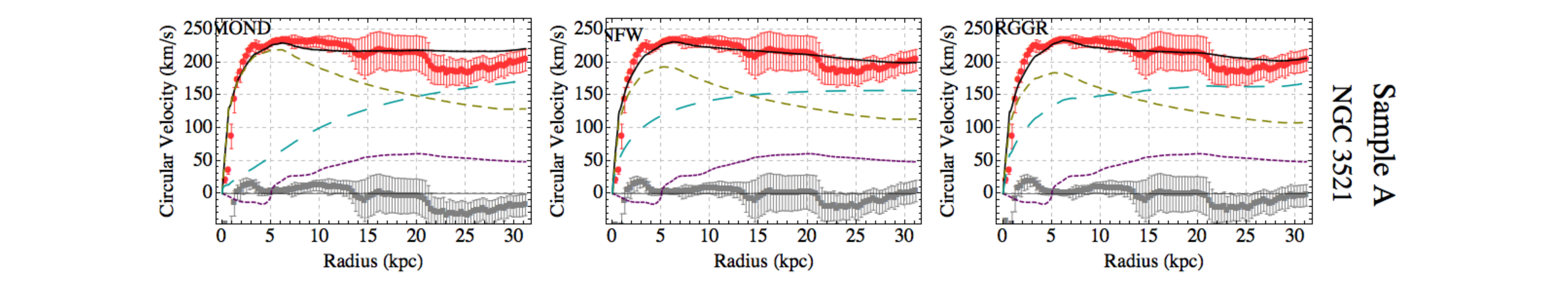}
\end{subfigure}
\begin{subfigure}
  \centering
  \opPlots{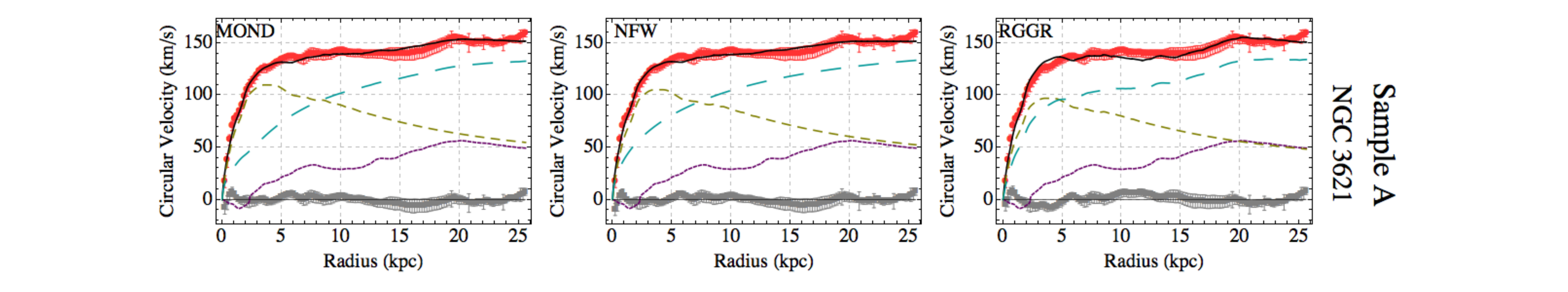}
\end{subfigure}
\begin{subfigure}
  \centering
  \opPlots{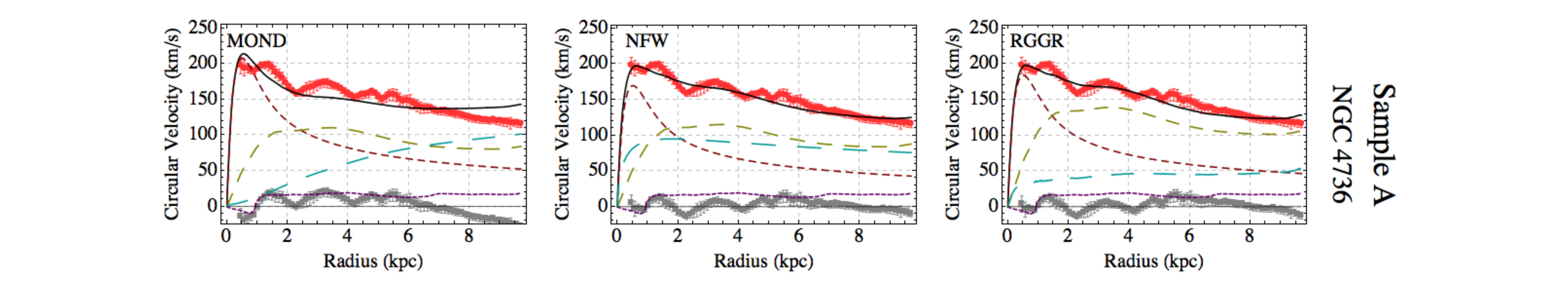}
\end{subfigure}
\begin{subfigure}
  \centering
  \opPlots{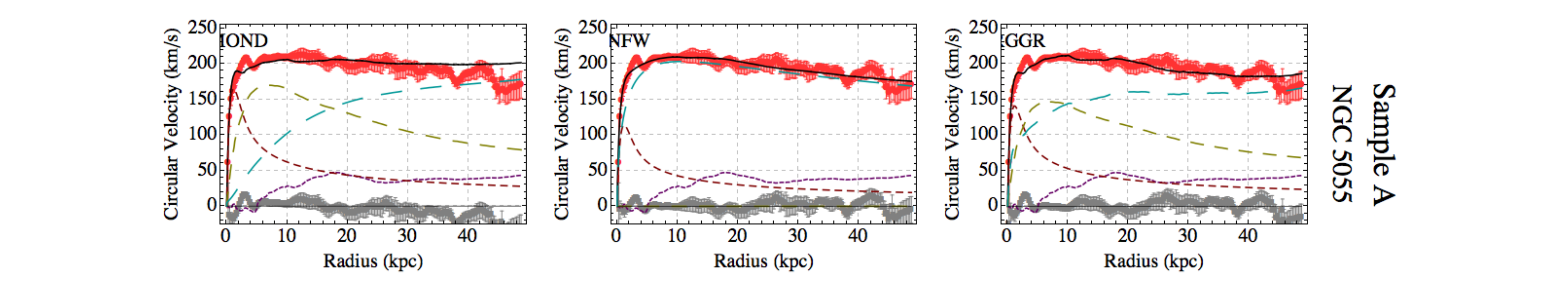}
\end{subfigure}
\begin{subfigure}
  \centering
  \opPlots{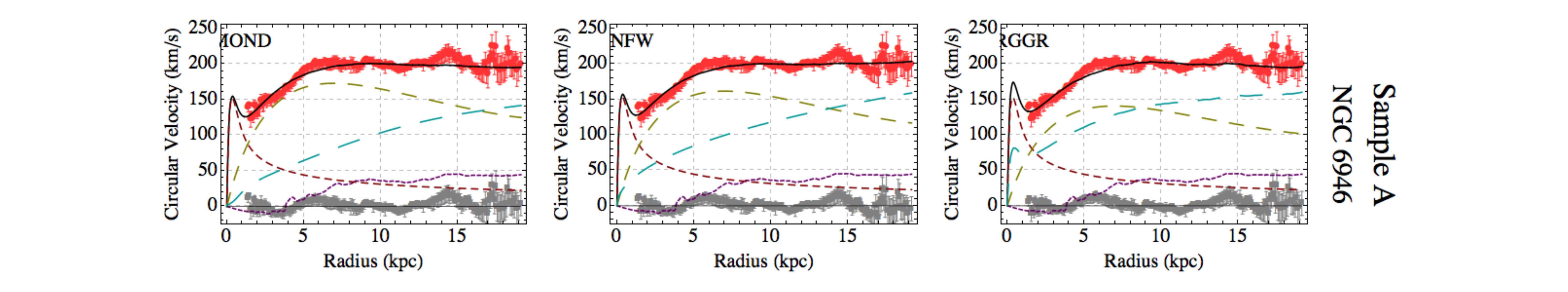}
\end{subfigure}
\begin{subfigure}
  \centering
  \opPlotsF{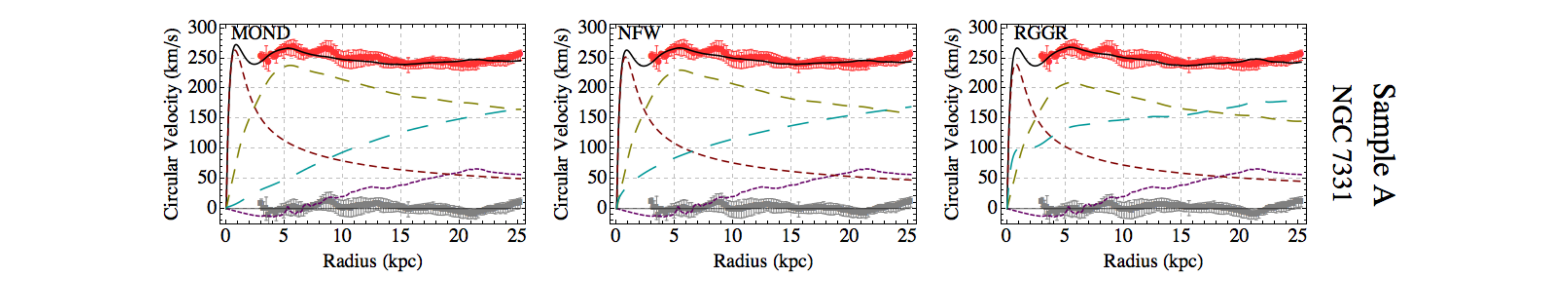}
\end{subfigure}
\contcaption{}
\end{figure*}

\begin{figure*}
\begin{subfigure}
  \centering
  \opPlots{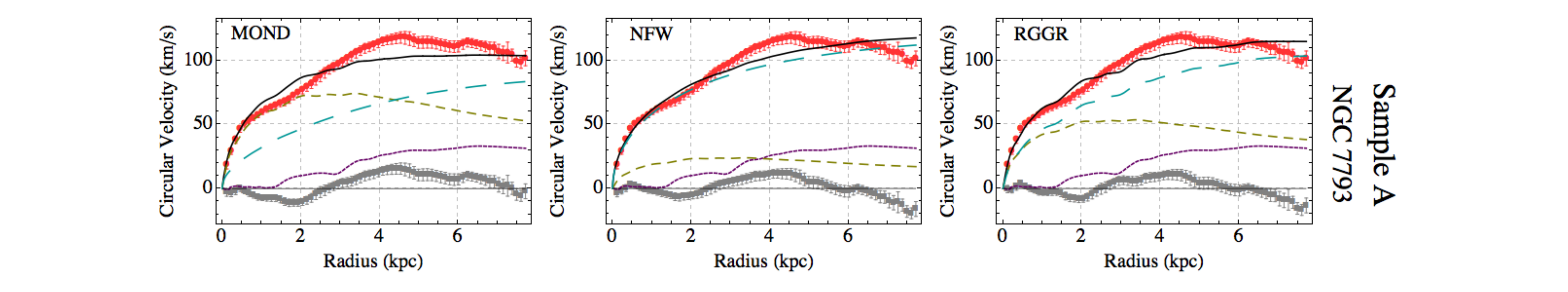}
\end{subfigure}
\begin{subfigure}
  \centering
  \opPlots{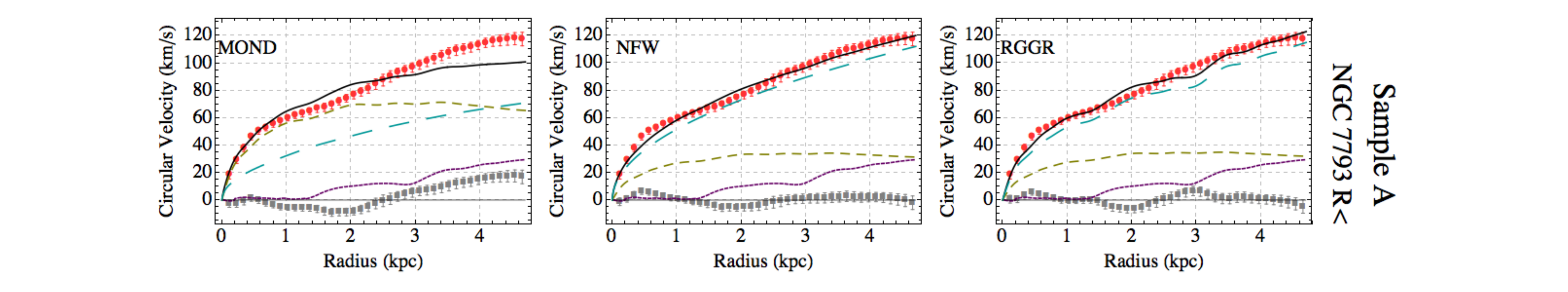}
\end{subfigure}
\begin{subfigure}
  \centering
  \opPlots{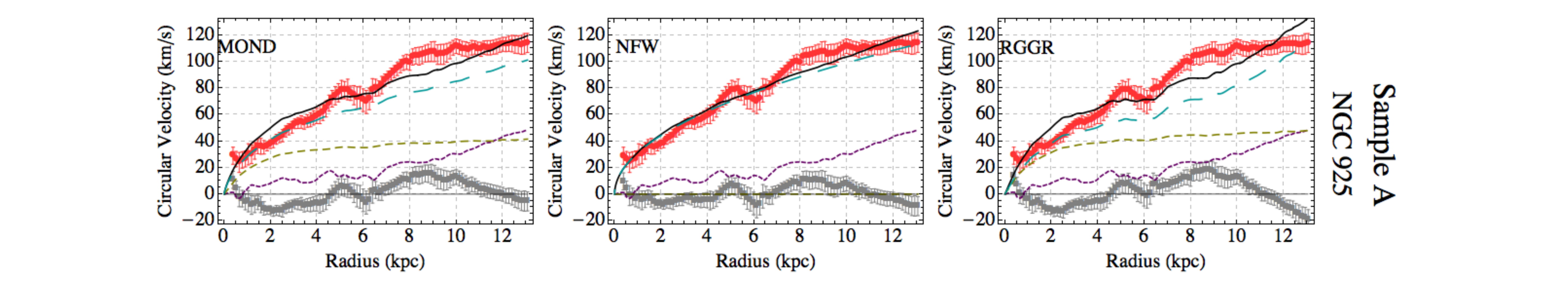}
\end{subfigure}
\begin{subfigure}
  \centering
  \opPlots{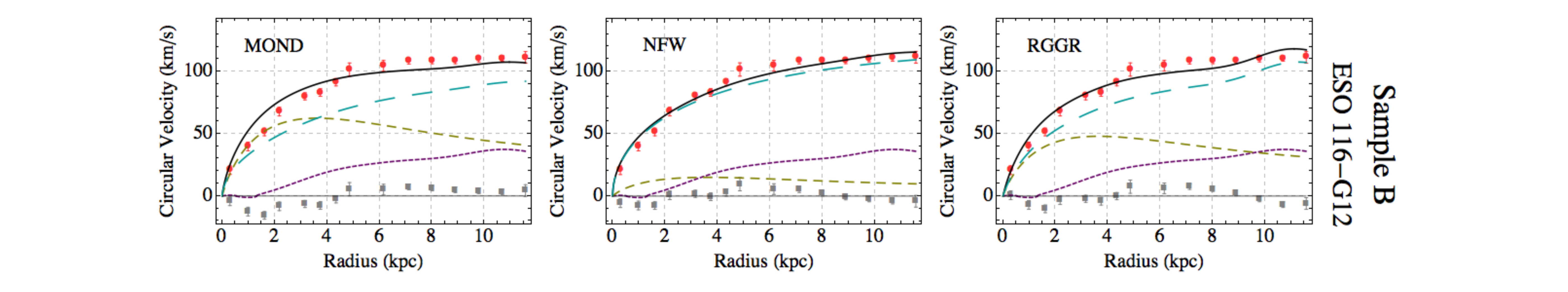}
  \label{fig:Gentileetal1}
\end{subfigure}%
\begin{subfigure}
  \centering
  \opPlots{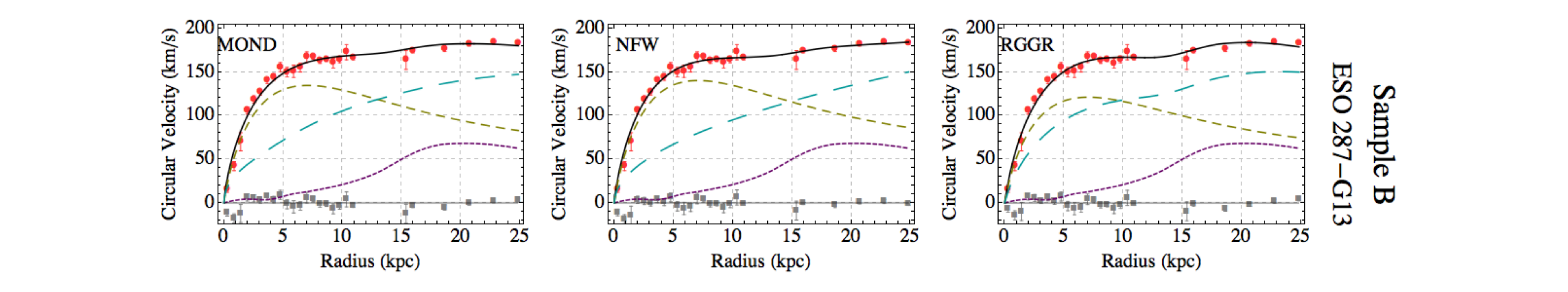}
  \label{fig:Gentileetal2}
\end{subfigure}%
\begin{subfigure}
  \centering
  \opPlots{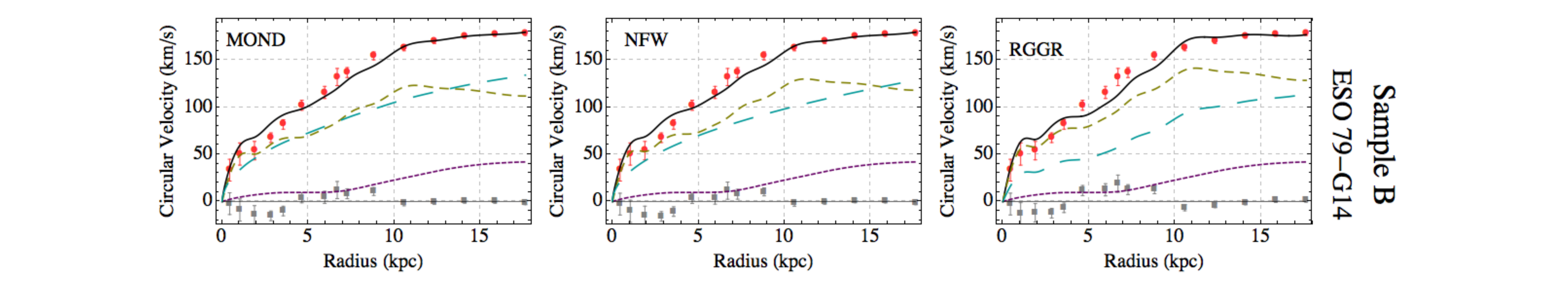}
  \label{fig:Gentileetal3}
\end{subfigure}%
\begin{subfigure}
  \centering
  \opPlots{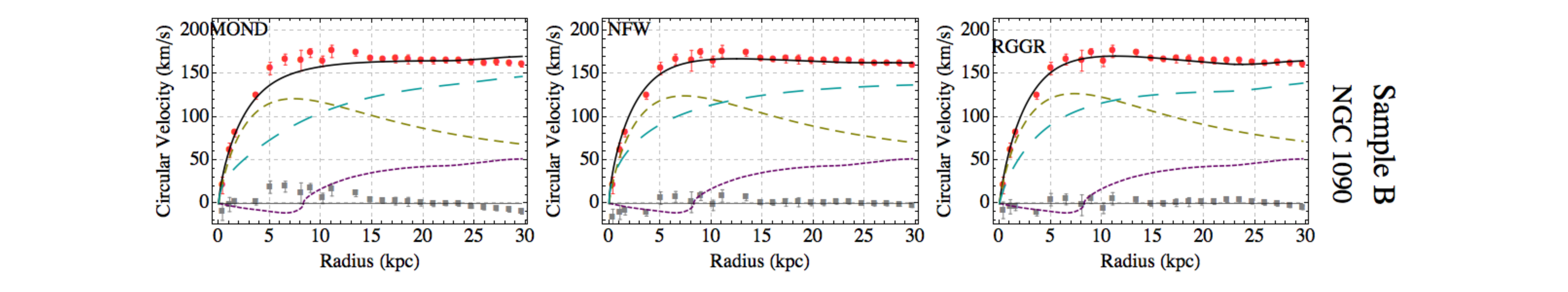}
  \label{fig:Gentileetal4}
\end{subfigure}%
\begin{subfigure}
  \centering
  \opPlotsF{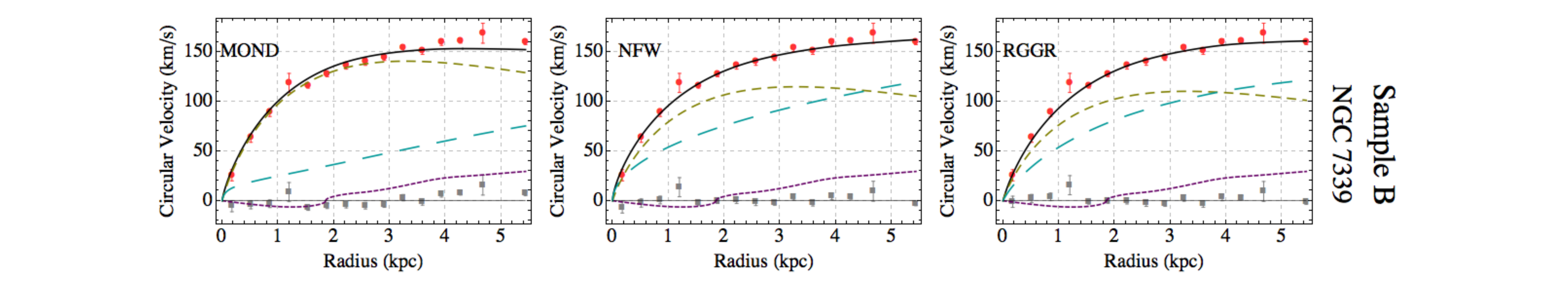}
  \label{fig:Gentileetal5}
\end{subfigure}%
\contcaption{} 
\end{figure*}

\begin{figure*}
\begin{subfigure}
  \centering
  \opPlots{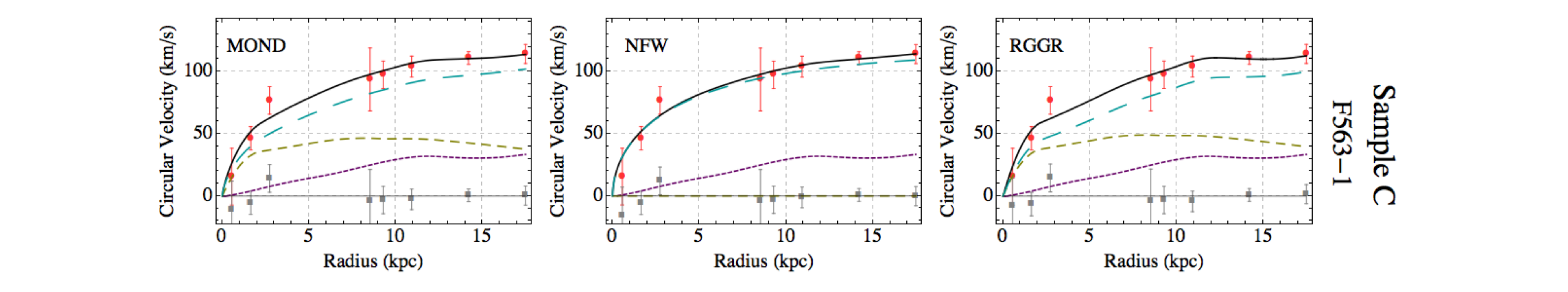}
\end{subfigure}%
\begin{subfigure}
  \centering
  \opPlots{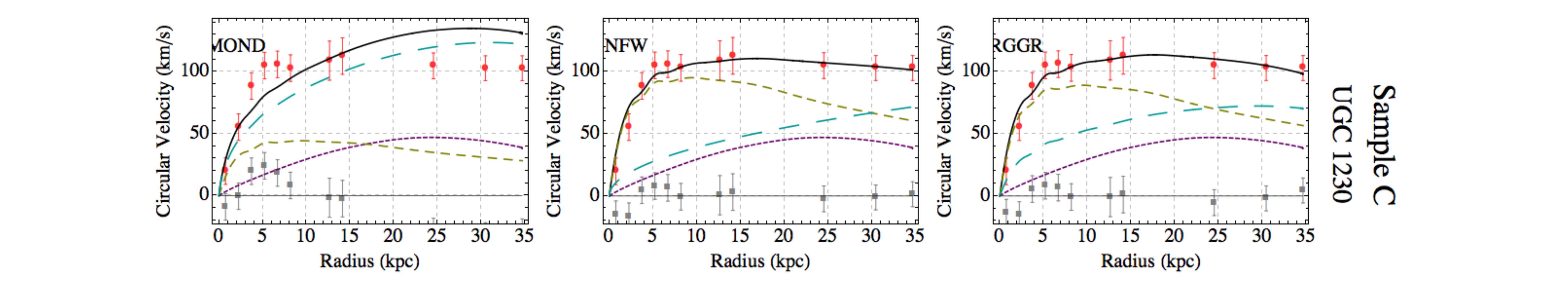}
\end{subfigure}%
\begin{subfigure}
  \centering
  \opPlots{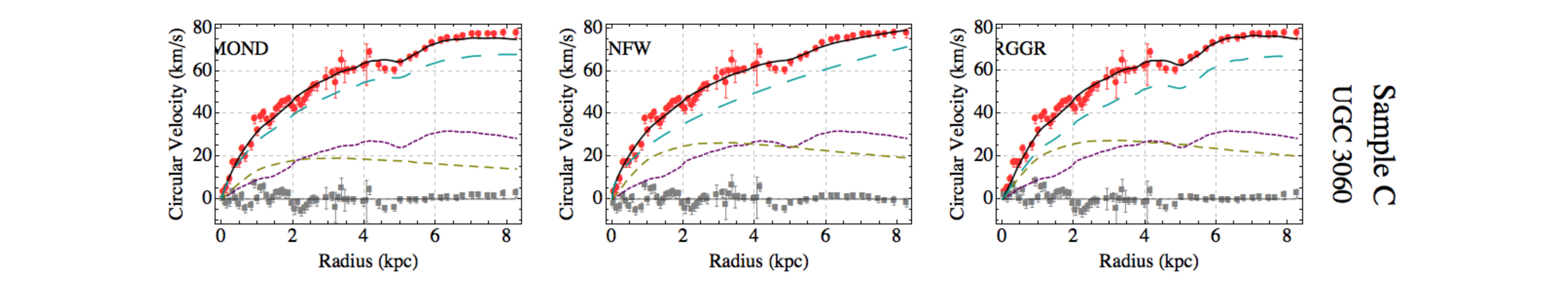}
\end{subfigure}%
\begin{subfigure}
  \centering
  \opPlots{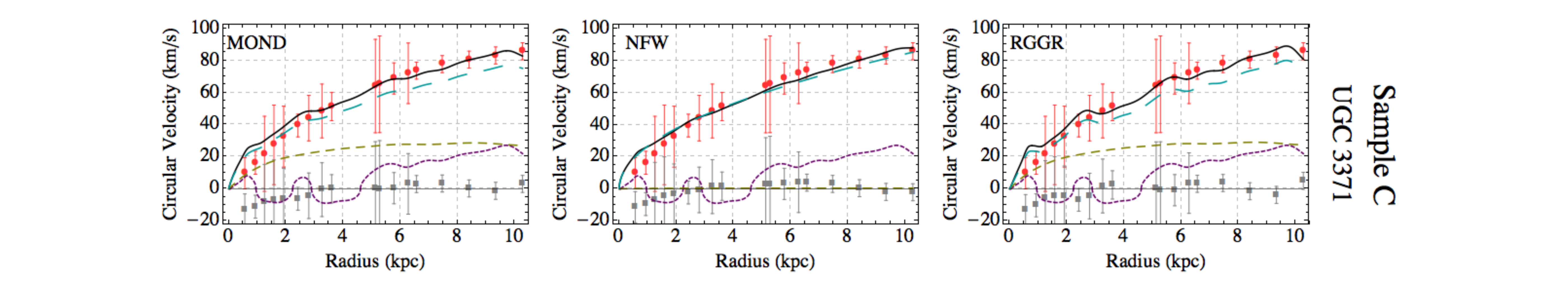}
\end{subfigure}%
\begin{subfigure}
  \centering
  \opPlots{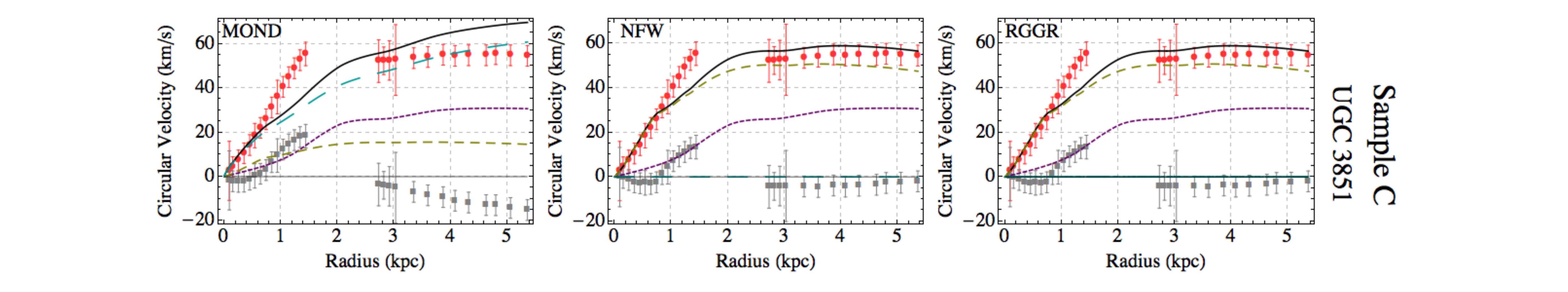}
\end{subfigure}%
\begin{subfigure}
  \centering
  \opPlots{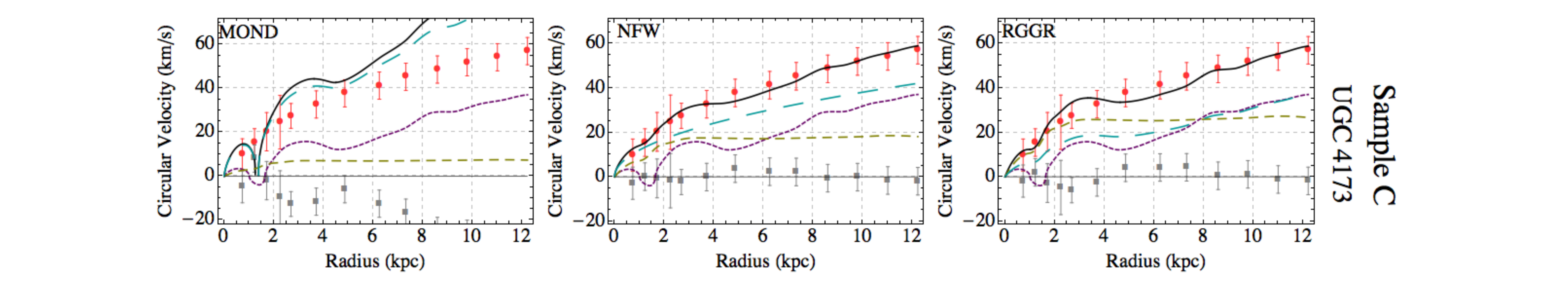}
\end{subfigure}%
\begin{subfigure}
  \centering
  \opPlots{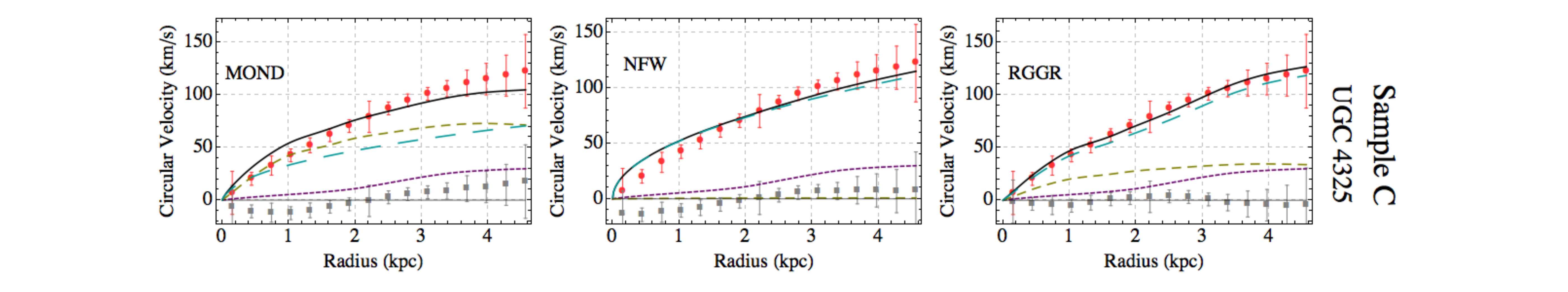}
\end{subfigure}%
\begin{subfigure}
  \centering
  \opPlotsF{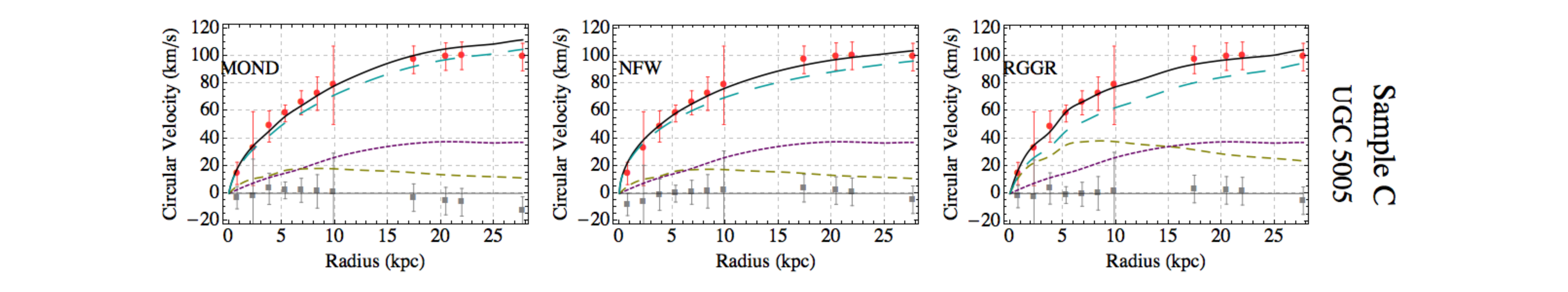}
\end{subfigure}%
\contcaption{}
\end{figure*}

\begin{figure*}
\begin{subfigure}
  \centering
  \opPlots{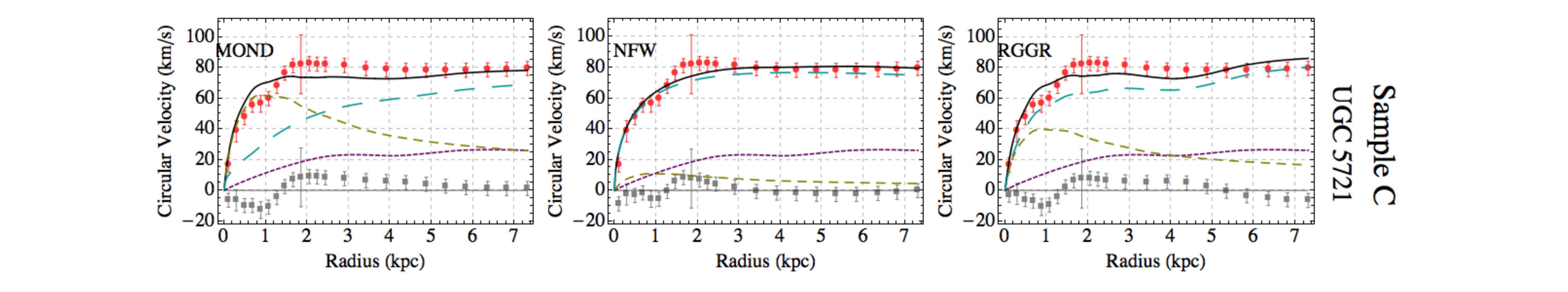}
\end{subfigure}%
\begin{subfigure}
  \centering
  \opPlots{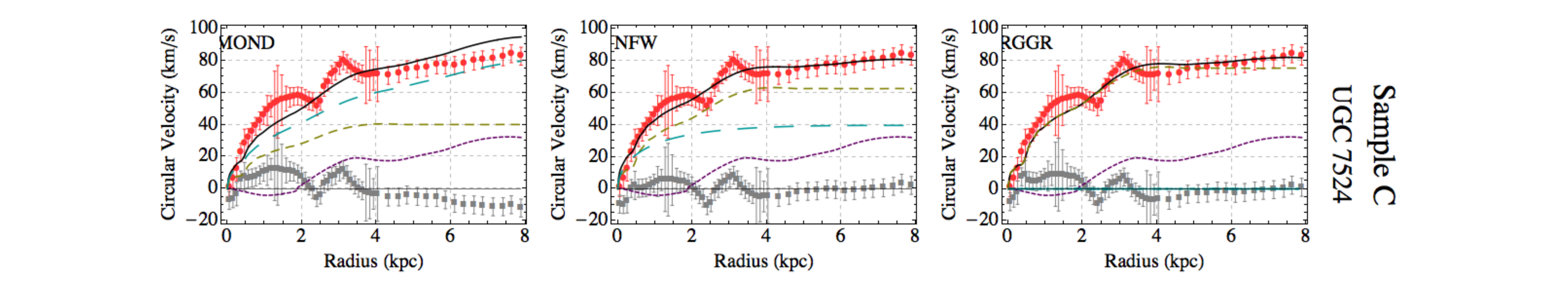}
\end{subfigure}%
\begin{subfigure}
  \centering
  \opPlots{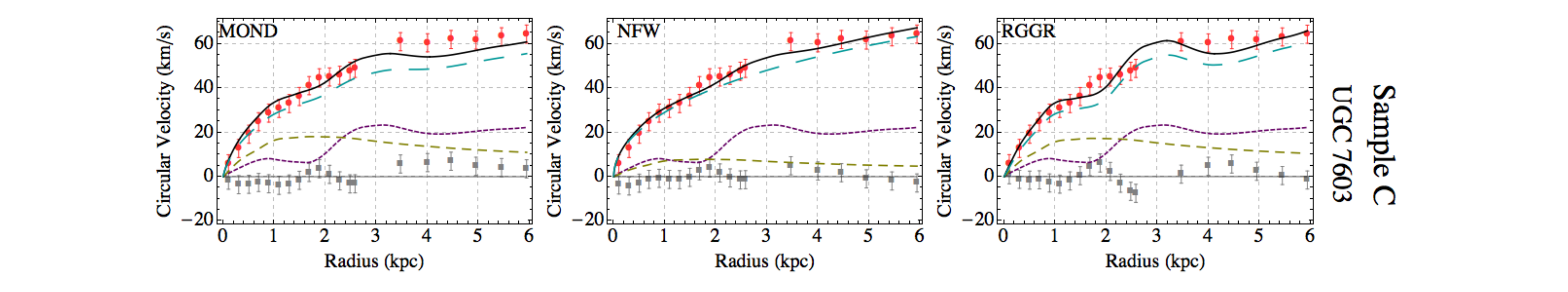}
\end{subfigure}%
\begin{subfigure}
  \centering
  \opPlots{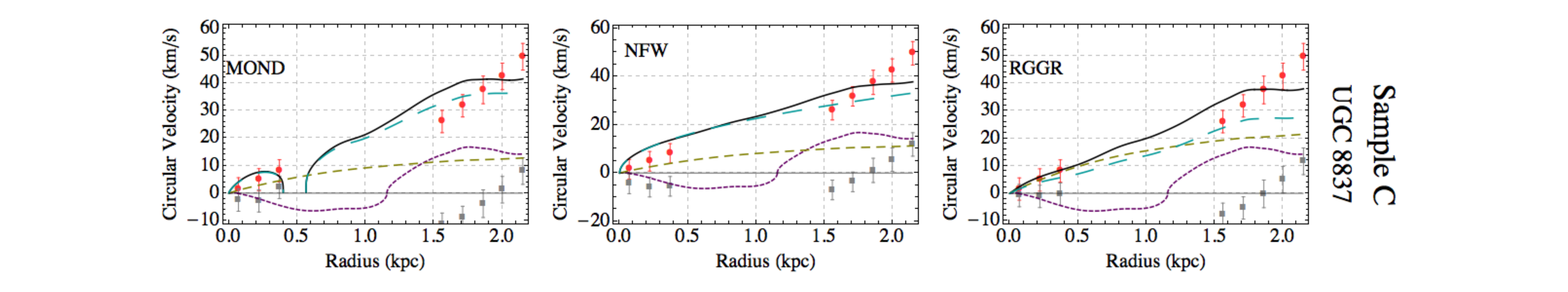}
\end{subfigure}%
\begin{subfigure}
  \centering
  \opPlots{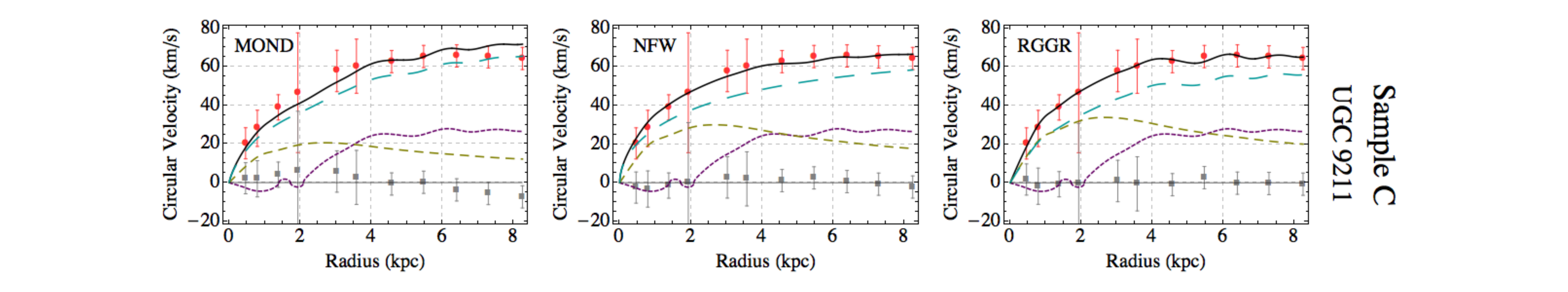}
\end{subfigure}%
\begin{subfigure}
  \centering
  \opPlots{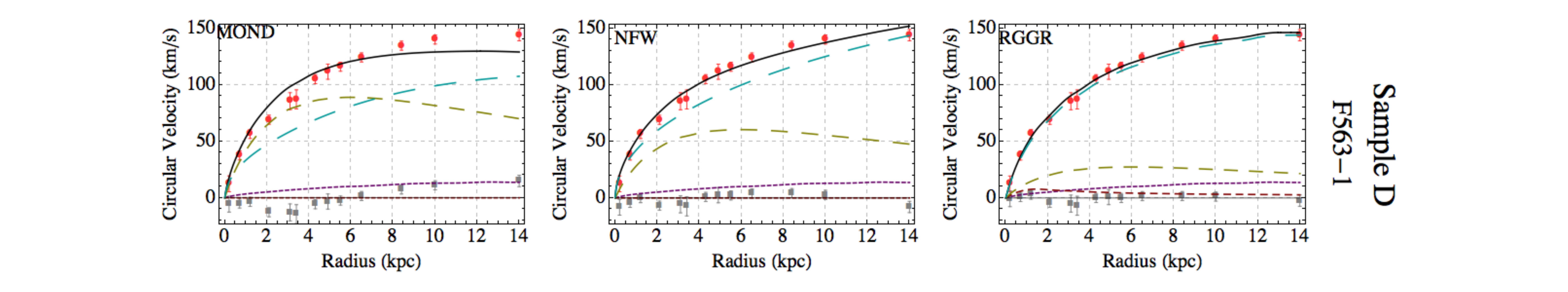}
  \label{fig:dbmr1}
\end{subfigure}%
\begin{subfigure}
  \centering
  \opPlots{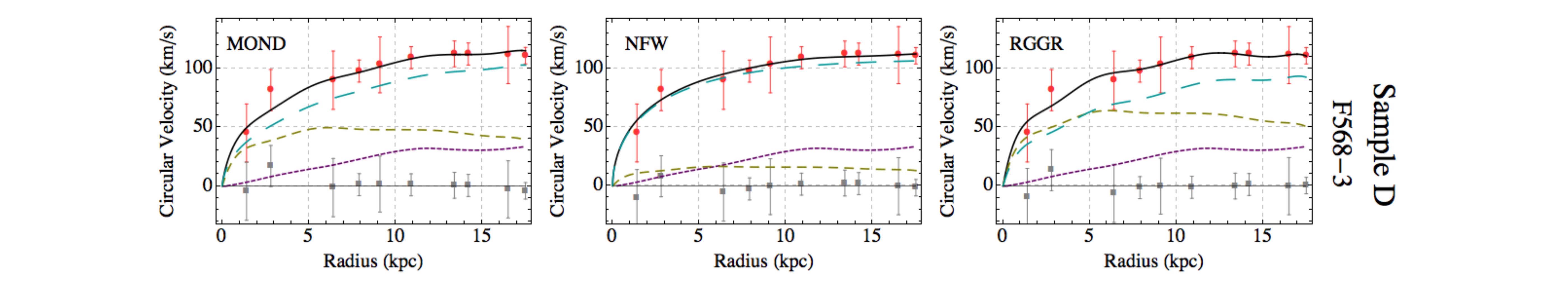}
  \label{fig:dbmr2}
\end{subfigure}
\begin{subfigure}
  \centering
  \opPlotsF{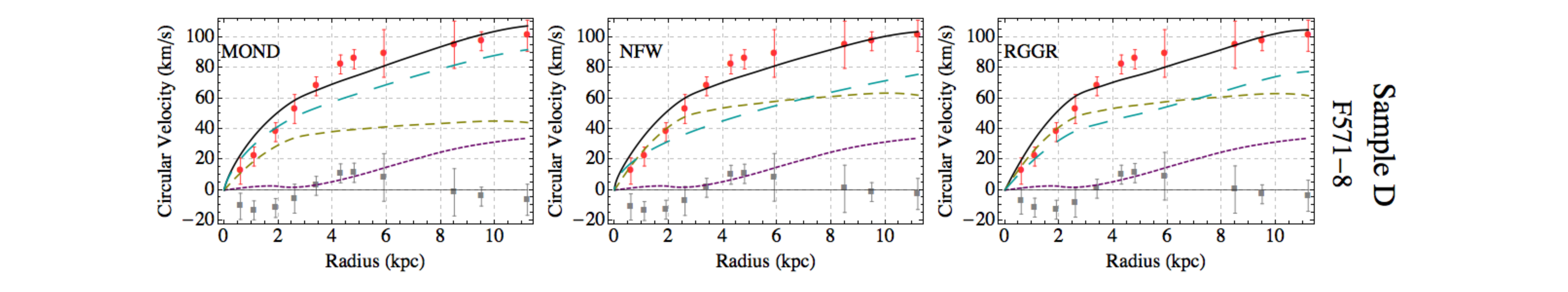}
  \label{fig:dbmr3}
\end{subfigure}
\contcaption{}
\end{figure*}

\begin{figure*}
\begin{subfigure}
  \centering
  \opPlots{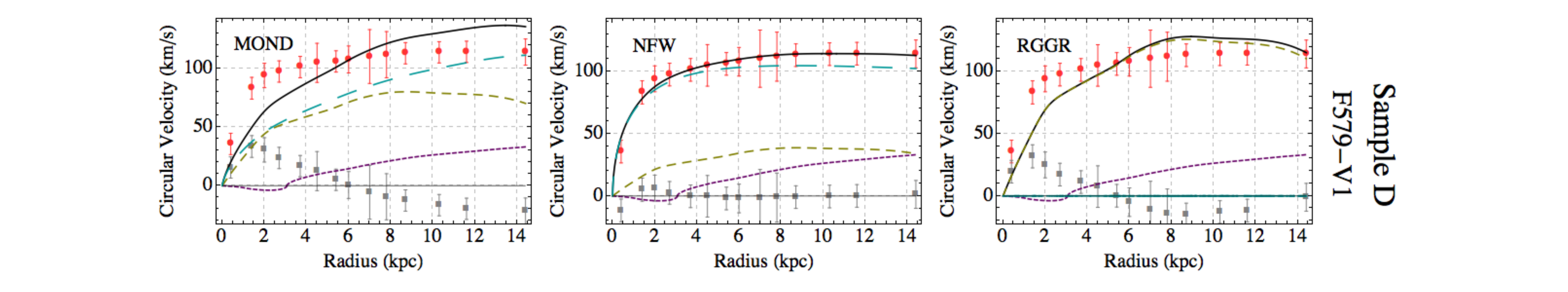}
  \label{fig:dbmr4}
\end{subfigure}
\begin{subfigure}
  \centering
  \opPlots{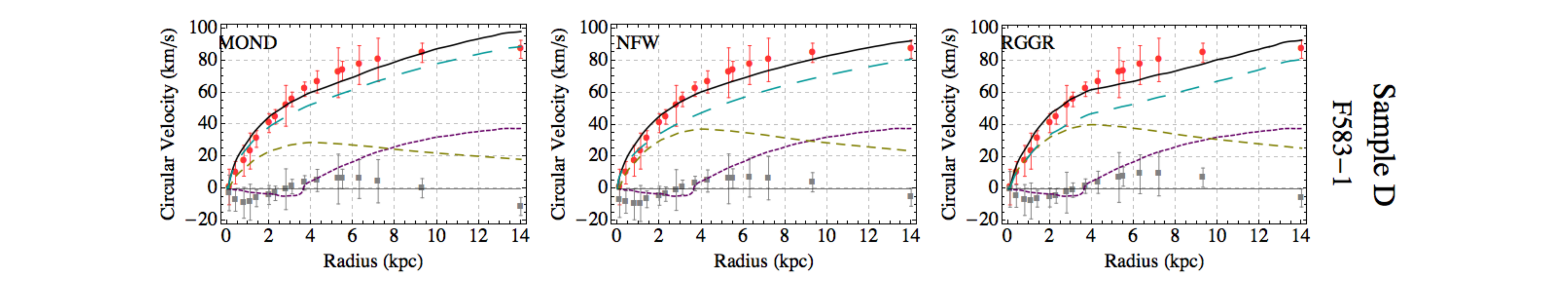}
  \label{fig:dbmr5}
\end{subfigure}
\begin{subfigure}
  \centering
  \opPlots{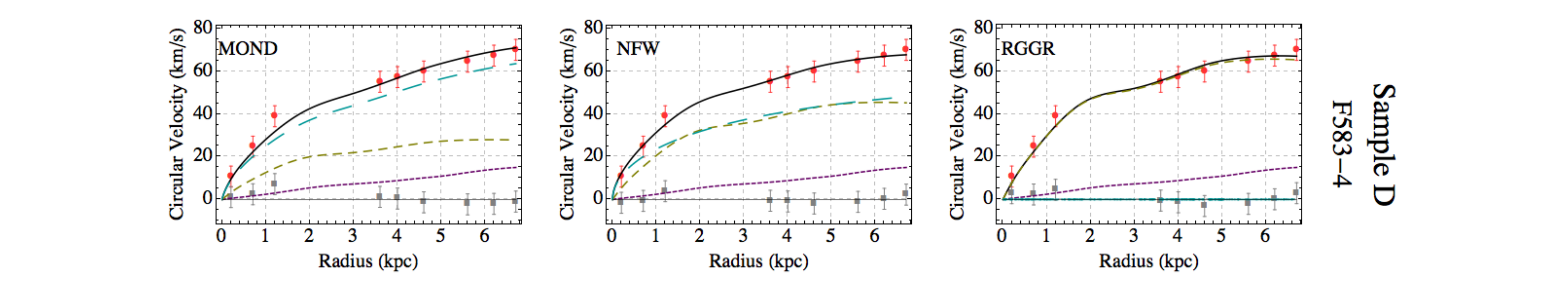}
  \label{fig:dbmr6}
\end{subfigure}
\begin{subfigure}
  \centering
  \opPlots{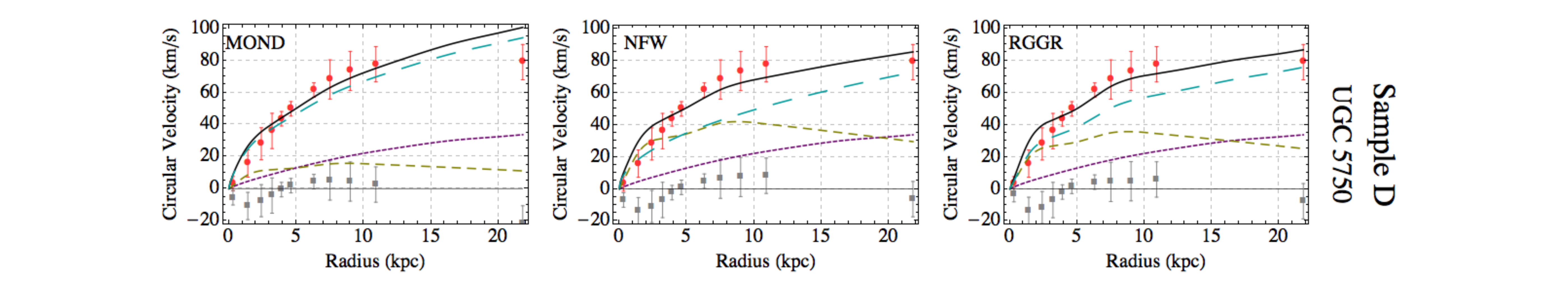}
  \label{fig:dbmr7}
\end{subfigure}
\begin{subfigure}
  \centering
  \opPlots{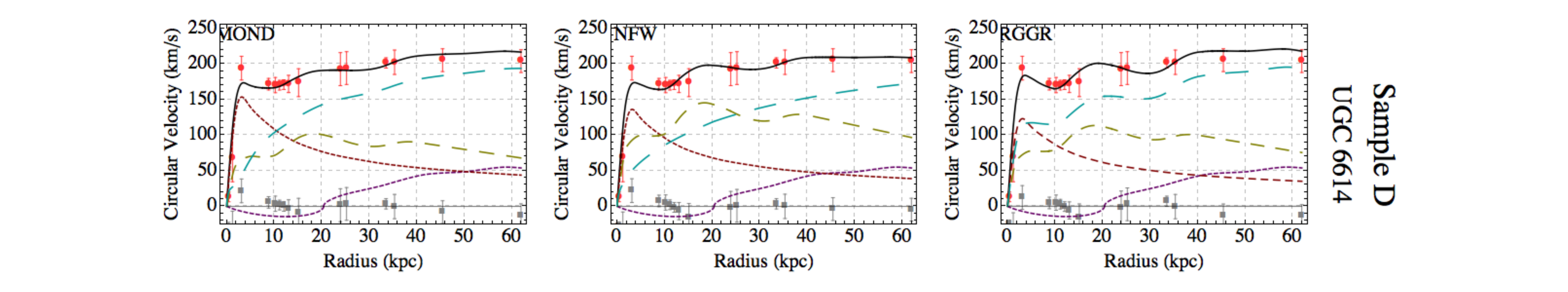}
  \label{fig:dbmr8}
\end{subfigure}
\begin{subfigure}
  \centering
  \opPlots{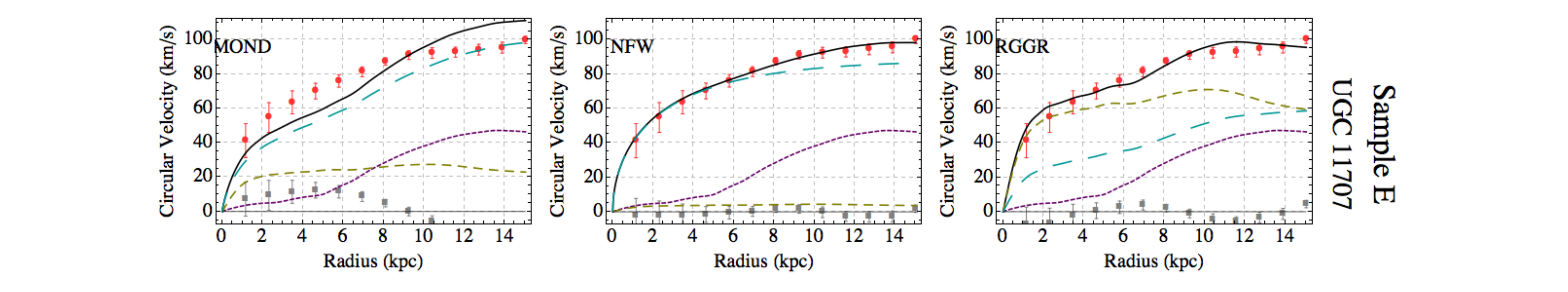}
\end{subfigure}
\begin{subfigure}
  \centering
  \opPlots{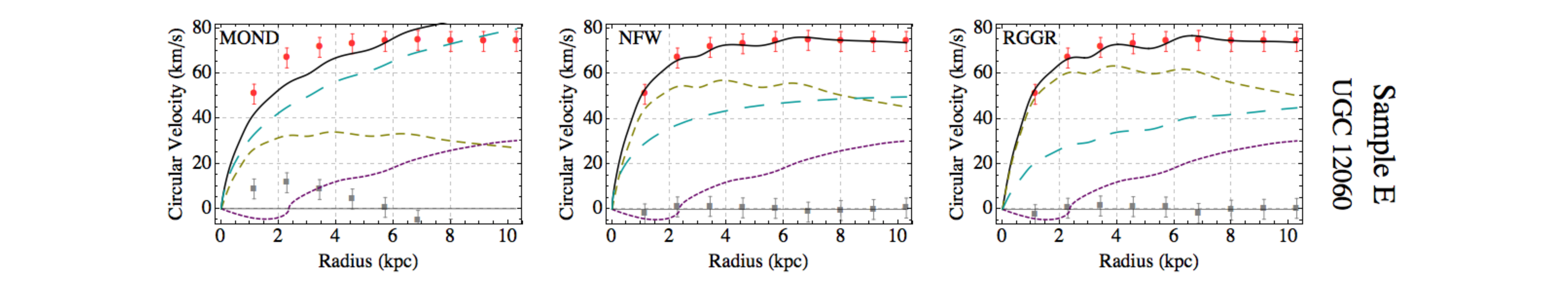}
\end{subfigure}
\begin{subfigure}
  \centering
  \opPlotsF{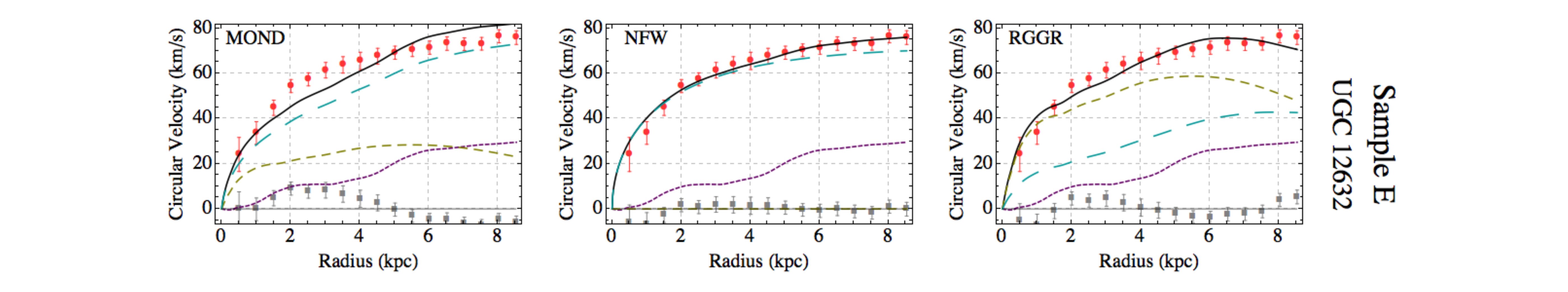}
\end{subfigure}%
\contcaption{}
\end{figure*}

\begin{figure*}
\begin{subfigure}
  \centering
  \opPlots{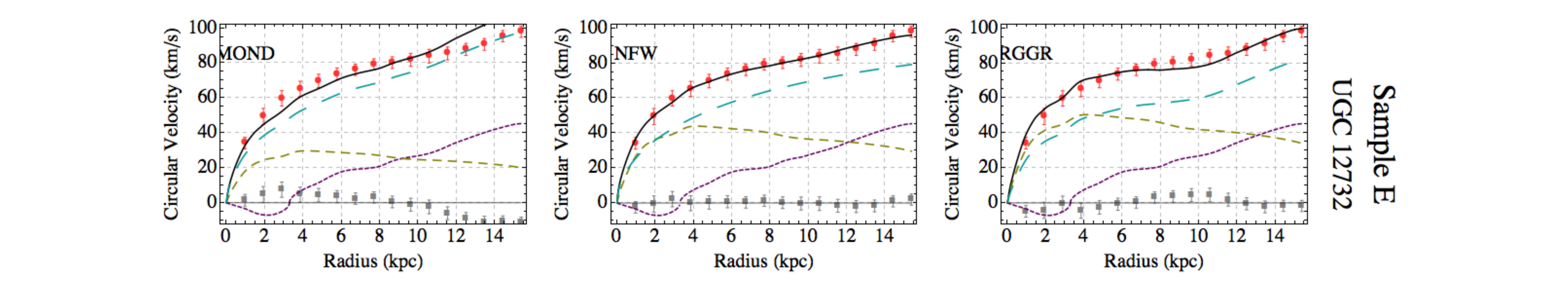}
\end{subfigure}
\begin{subfigure}
  \centering
  \opPlots{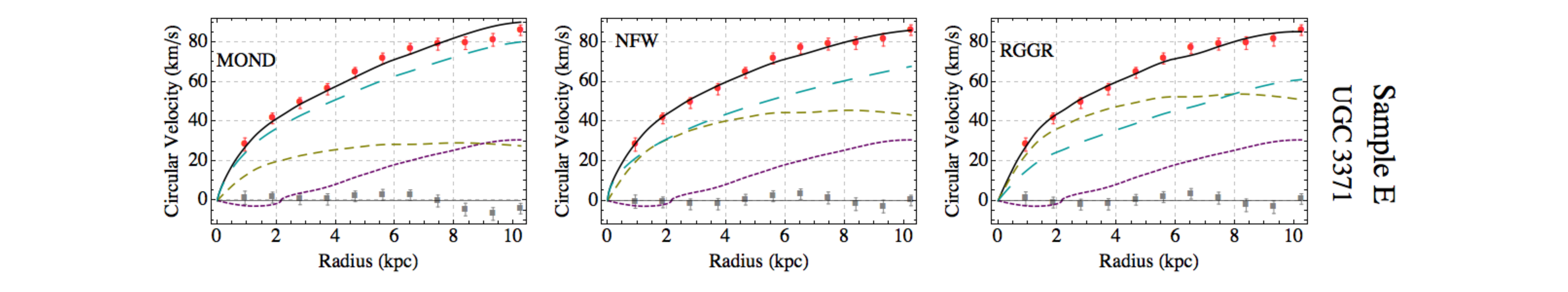}
\end{subfigure}
\begin{subfigure}
  \centering
  \opPlots{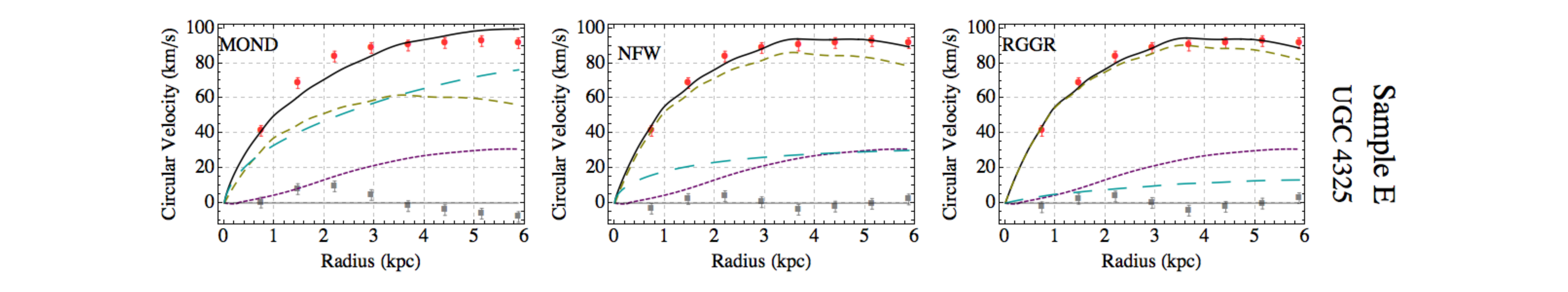}
\end{subfigure}
\begin{subfigure}
  \centering
  \opPlots{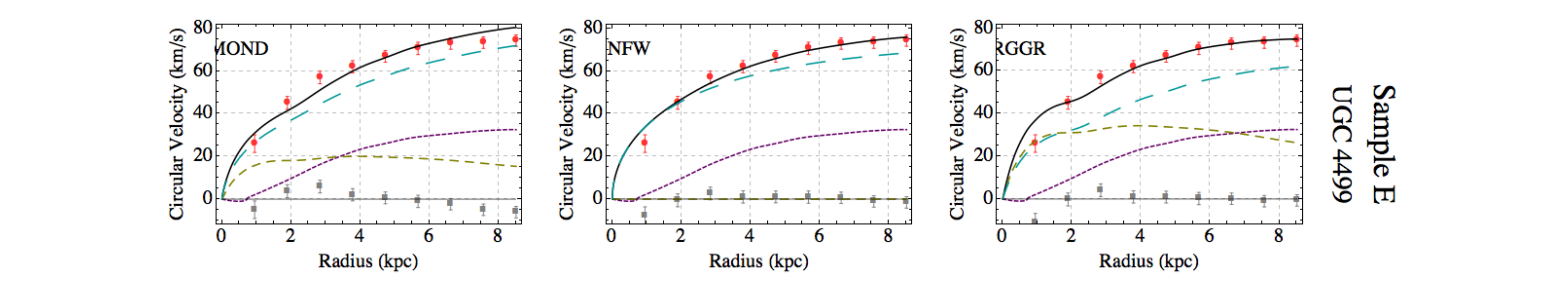}
\end{subfigure}
\begin{subfigure}
  \centering
  \opPlots{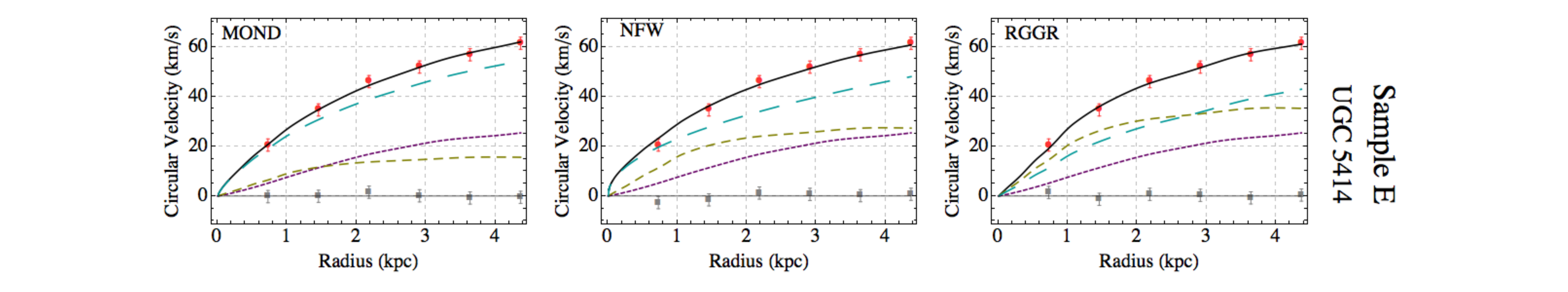}
\end{subfigure}
\begin{subfigure}
  \centering
  \opPlots{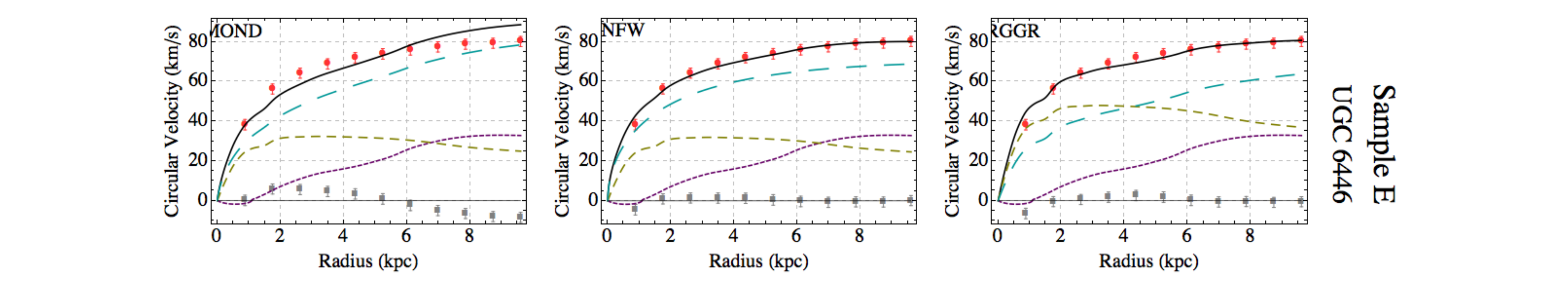}
\end{subfigure}
\begin{subfigure}
  \centering
  \opPlots{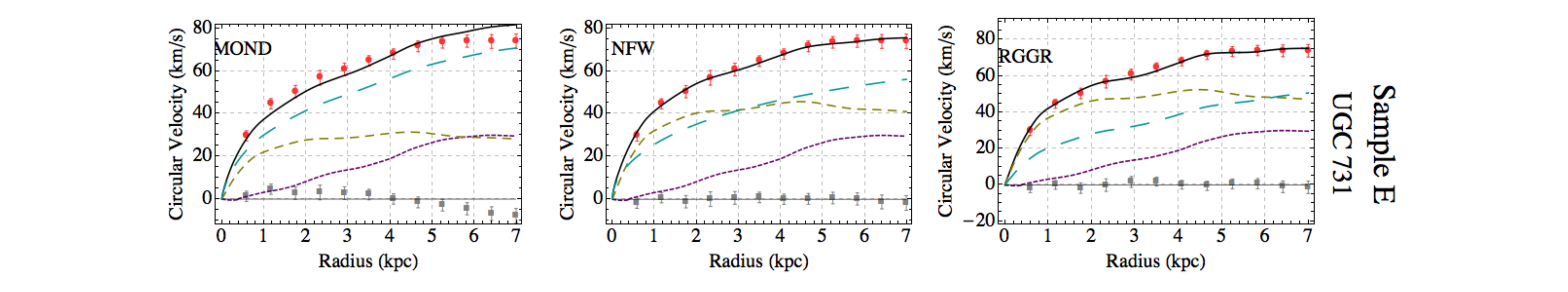}
\end{subfigure}%
\begin{subfigure}
  \centering
  \opPlotsF{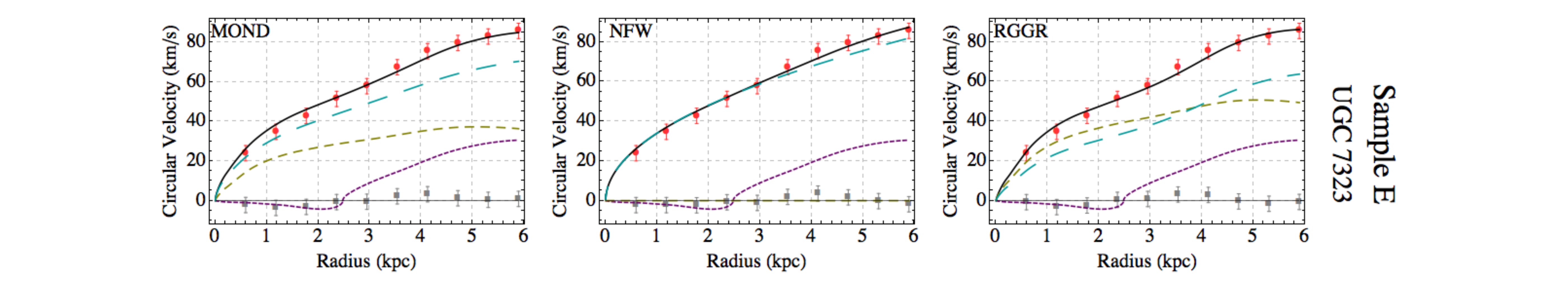}
\end{subfigure}
\contcaption{}
\end{figure*}

\begin{figure*}
\begin{subfigure}
  \centering
  \opPlots{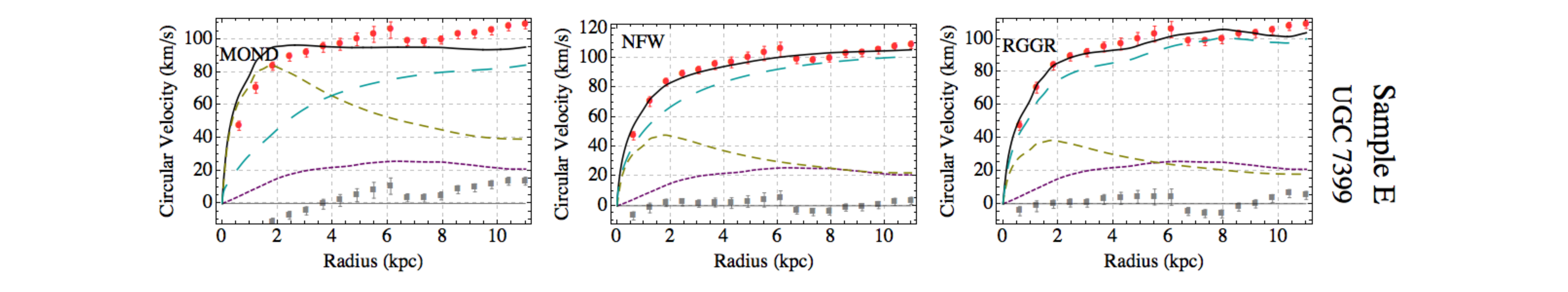}
\end{subfigure}
\begin{subfigure}
  \centering
  \opPlots{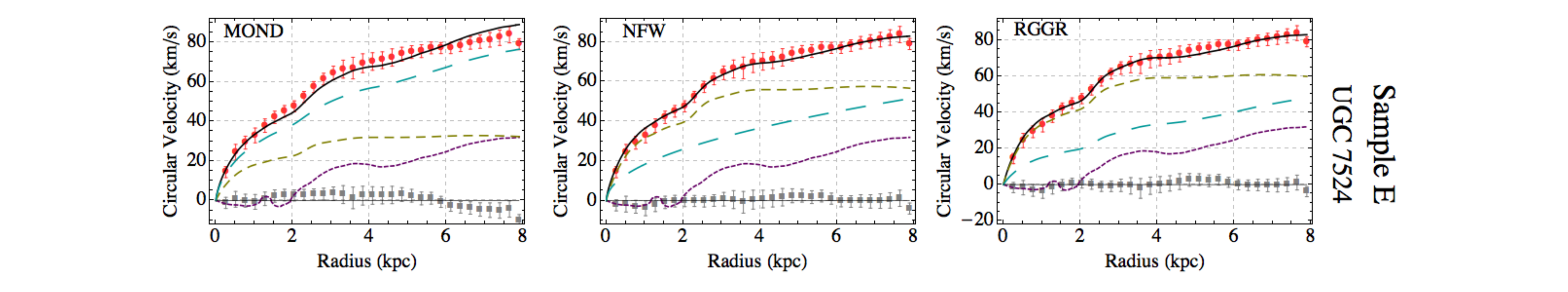}
\end{subfigure}%
\begin{subfigure}
  \centering
  \opPlots{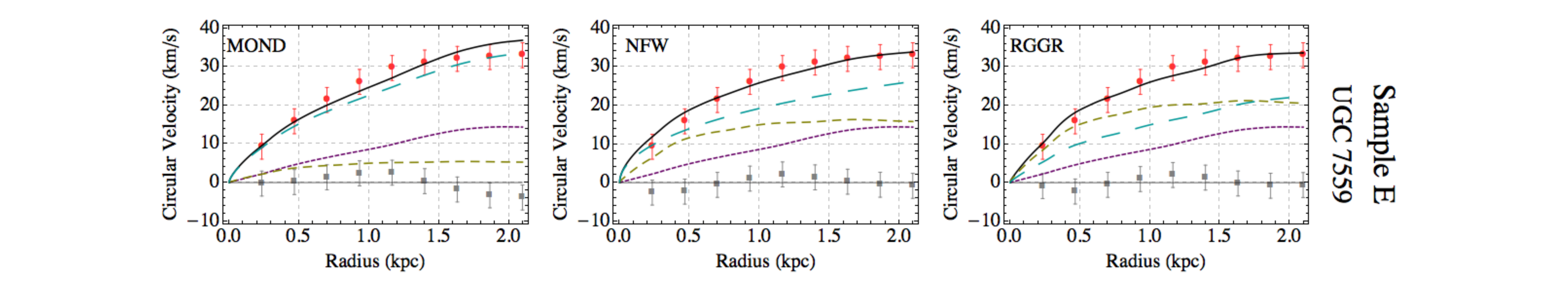}
\end{subfigure}
\begin{subfigure}
  \centering
  \opPlots{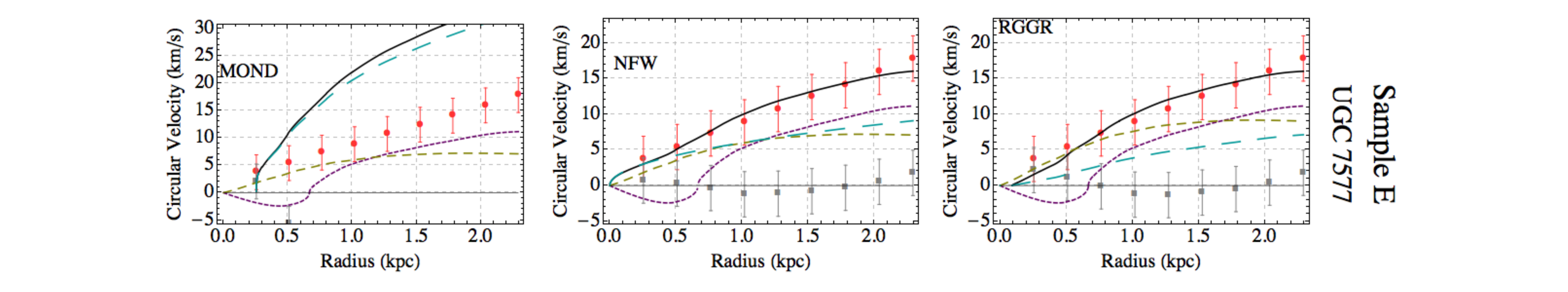}
\end{subfigure}
\begin{subfigure}
  \centering
  \opPlots{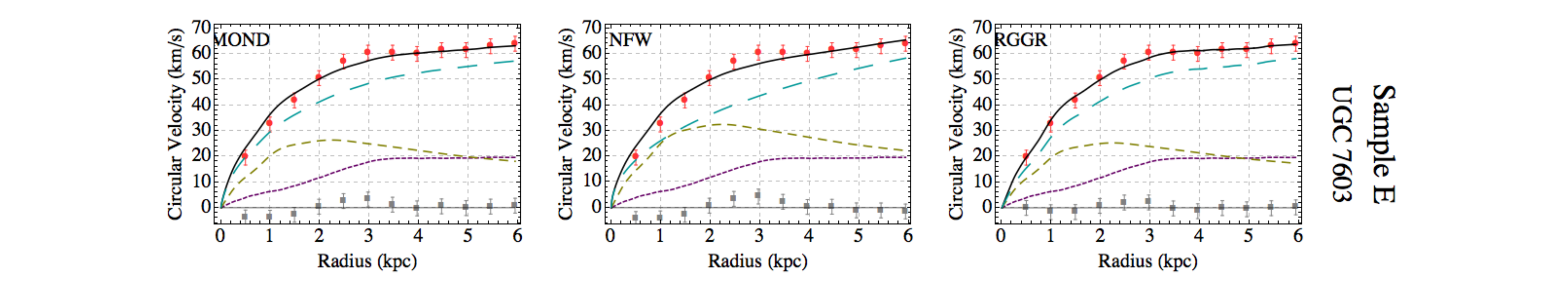}
\end{subfigure}
\begin{subfigure}
  \centering
  \opPlots{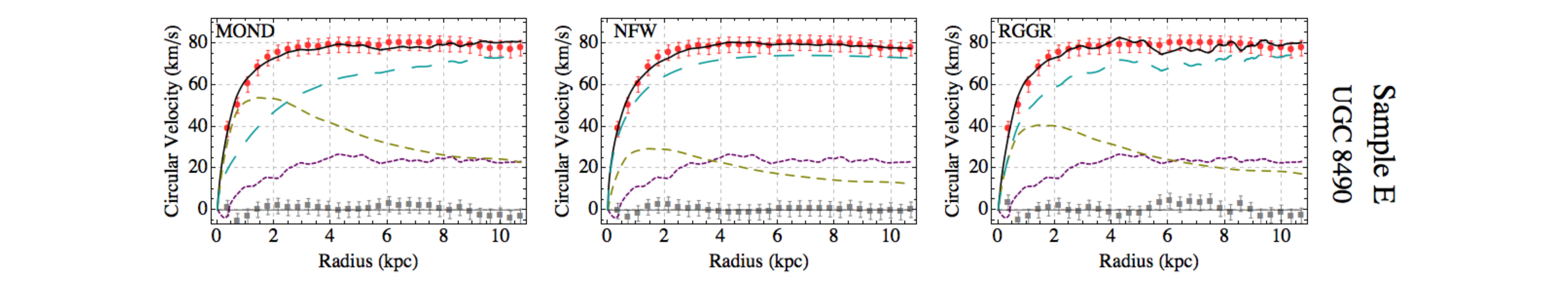}
\end{subfigure}
\begin{subfigure}
  \centering
  \opPlotsF{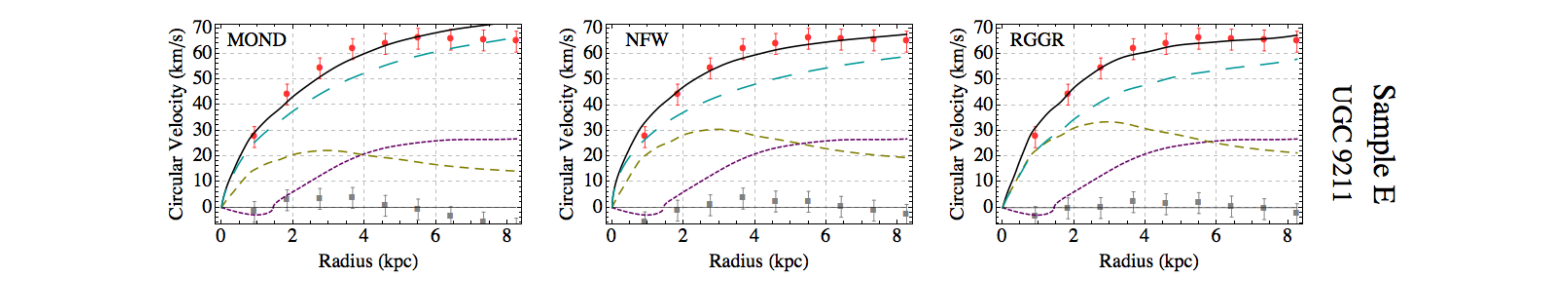}
\end{subfigure}
\contcaption{}
\end{figure*}

\label{lastpage}
\end{document}